%% file: main.tex
\documentclass[acmtog,screen]{acmart}
\usepackage{multirow}
\usepackage{enumitem}
\usepackage{prettyref}

\usepackage[labelfont=bf,format=plain,singlelinecheck=false]{caption}
\usepackage{subcaption}
\usepackage{bm,microtype}
\setcopyright{acmlicensed}
\acmJournal{TOG}
\acmYear{2025} \acmVolume{44} \acmNumber{6} \acmArticle{1} \acmMonth{12}\acmDOI{10.1145/3763295}

\newcommand{\rev}[1]{\textcolor[RGB]{255, 0, 0}{{#1}}}
\definecolor{myblue}{RGB}{31, 119, 180}
\definecolor{myorange}{RGB}{255, 127, 14}
\definecolor{mygreen}{RGB}{44, 160, 44}
\definecolor{myred}{RGB}{214, 39, 40}
\usepackage{soul}
\AtBeginDocument{%
  }

\acmJournal{TOG}

\usepackage{tikz}
\usetikzlibrary{angles, calc}
\tikzset{
  dot/.style={
    circle, fill=black, inner sep=1pt, outer sep=0pt
  },
  dot label/.style={
    circle, inner sep=0pt, outer sep=1pt
  },
  pics/right angle/.append style={
    /tikz/draw, /tikz/angle radius=5pt
  }
}
\usepackage{xcolor}

\definecolor{mycard}{RGB}{255,192,0}
\definecolor{mytex}{RGB}{204,0,0}
\definecolor{mypsnr}{RGB}{255,102,0}
\definecolor{mylpips}{RGB}{61,133,198}
\definecolor{mycoverage}{RGB}{106,168,79}

\acmSubmissionID{1299}

\usepackage[noend]{algpseudocode}

\def\BState{\State\hskip-\ALG@thistlm}
\makeatother

\newcommand{\algorithmicforeach}{\textbf{foreach}}

\algdef{SE}[FOREACH]{ForEach}{EndForEach}[1]{\algorithmicforeach\ #1\ \algorithmicdo}{\algorithmicend\ \algorithmicforeach}%
\algtext*{EndForEach}

\usepackage{prettyref}
\newrefformat{fig}{Figure~\ref{#1}}
\newrefformat{par}{Section~\ref{#1}}
\newrefformat{appen}{Appendix~\ref{#1}}
\newrefformat{sec}{Section~\ref{#1}}
\newrefformat{sub}{Section~\ref{#1}}
\newrefformat{table}{Table~\ref{#1}}
\newrefformat{ass}{Assumption~\ref{#1}}
\newrefformat{alg}{Algorithm~\ref{#1}}
\newrefformat{def}{Definition~\ref{#1}}
\newrefformat{thm}{Theorem~\ref{#1}}
\newrefformat{cor}{Corollary~\ref{#1}}
\newrefformat{lem}{Lemma~\ref{#1}}
\newrefformat{step}{Step~\ref{#1}}
\newrefformat{ln}{Line~\ref{#1}}
\newrefformat{rem}{Remark~\ref{#1}}
\newrefformat{eq}{Equation~\ref{#1}}
\newrefformat{pb}{Problem~\ref{#1}}
\newrefformat{it}{Item~\ref{#1}}
\newrefformat{te}{Term~\ref{#1}}
\def\Eqref Eq:#1:{\eqref{eq:#1}}
\newrefformat{Eq}{Equation~\Eqref#1:}

\usepackage{rotating}

\newcommand{\bb}{\mathbf{b}}

\newcommand{\pp}{\mathbf{p}}

\renewcommand{\ss}{\mathbf{s}}

\renewcommand{\tt}{\mathbf{t}}

\newcommand{\nn}{\mathbf{n}}

\newcommand\restr[2]{{\left.\kern-\nulldelimiterspace{}#1\right|_{#2}}}

\usepackage{algorithm} 
\usepackage{algpseudocode} 
\usepackage{svg}
\usepackage{xcolor}
\usepackage{wrapfig}
\usepackage[export]{adjustbox}
\usepackage{colortbl}

\usepackage{diagbox} 
\usepackage{multirow} 
\begin{document}

\title{Auto Hair Card Extraction for Smooth Hair with Differentiable Rendering}

\author{Zhongtian Zheng}
\orcid{0009-0009-4714-1760}
\email{zhengzhongtian@pku.edu.cn}
\affiliation{%
  \institution{LIGHTSPEED}
    \city{Shenzhen}
  \state{Guangzhou}
  \country{China}
}

\author{Tao Huang}
\orcid{0009-0002-3458-0851}
\email{tao_huang@ucsb.edu}
\affiliation{%
  \institution{LIGHTSPEED}
    \city{Los Angeles}
  \state{CA}
  \country{USA}
}

\author{Haozhe Su}
\orcid{0009-0002-8534-8964}
\email{dyhard0520@gmail.com}
\affiliation{%
  \institution{LIGHTSPEED}
    \city{Los Angeles}
  \state{CA}
  \country{USA}
}

\author{Xueqi Ma}
\orcid{0009-0004-0203-8501}
\email{qixuemaa@gmail.com}
\affiliation{%
  \institution{LIGHTSPEED}
    \city{Shenzhen}
  \state{Guangzhou}
  \country{China}
}

\author{Yuefan Shen}
\orcid{0000-0002-6049-7966}
\email{yuefanshen@outlook.com}
\affiliation{%
  \institution{LIGHTSPEED}
    \city{Shenzhen}
  \state{Guangzhou}
  \country{China}
}

\author{Tongtong Wang}
\orcid{0009-0005-6585-3009}
\email{wangtong923@gmail.com}
\affiliation{%
  \institution{LIGHTSPEED}
    \city{Shenzhen}
  \state{Guangzhou}
  \country{China}
}

\author{Yin Yang}
\orcid{0000-0001-7645-5931}
\email{yangzzzy@gmail.com}
\affiliation{%
  \institution{University of Utah}
    \city{Salt Lake City}
  \state{UT}
  \country{USA}
}

\author{Xifeng Gao}
\orcid{0000-0003-0829-7075}
\email{gxfxisha@gmail.com}
\affiliation{%
  \institution{LIGHTSPEED}
    \city{Seattle}
  \state{WA}
  \country{USA}
}

\author{Zherong Pan}
\orcid{0000-0001-9348-526X}
\email{zherong.pan.usa@gmail.edu}
\affiliation{%
  \institution{LIGHTSPEED}
    \city{Seattle}
  \state{WA}
  \country{USA}
}

\author{Kui Wu}
\orcid{0000-0003-3326-7943}
\email{walker.kui.wu@gmail.com}
\affiliation{%
  \institution{LIGHTSPEED}
    \city{Los Angeles}
  \state{CA}
  \country{USA}
}

\begin{abstract}
Hair cards remain a widely used representation for hair modeling in real-time applications, offering a practical trade-off between visual fidelity, memory usage, and performance. However, generating high-quality hair card models remains a challenging and labor-intensive task. This work presents an automated pipeline for converting strand-based hair models into hair card models with a limited number of cards and textures while preserving the hairstyle appearance. 
Our key idea is a novel differentiable representation where each strand is encoded as a projected 2D curve in the texture space, which enables end-to-end optimization with differentiable rendering while respecting the structures of the hair geometry. Based on this representation, we develop a novel algorithm pipeline, where we first cluster hair strands into initial hair cards and project the strands into the texture space. We then conduct a two-stage optimization, where our first stage optimizes the orientation of each hair card separately, and after strand projection, our second stage conducts joint optimization over the entire hair card model for fine-tuning.
Our method is evaluated on a range of hairstyles, including straight, wavy, curly, and coily hair. To capture the appearance of short or coily hair, our method comes with support for hair caps and cross-card. 
\end{abstract}

\ccsdesc[500]{Computing methodologies~Shape modeling}

%
%

\keywords{Hair; hair card; differentiable rendering}

\begin{teaserfigure}
\newcommand{\figcap}[1]{\begin{minipage}{0.13\linewidth}\centering#1\end{minipage}}
\centering
\includegraphics[trim =0 0 80 80, clip, width=0.15\linewidth]{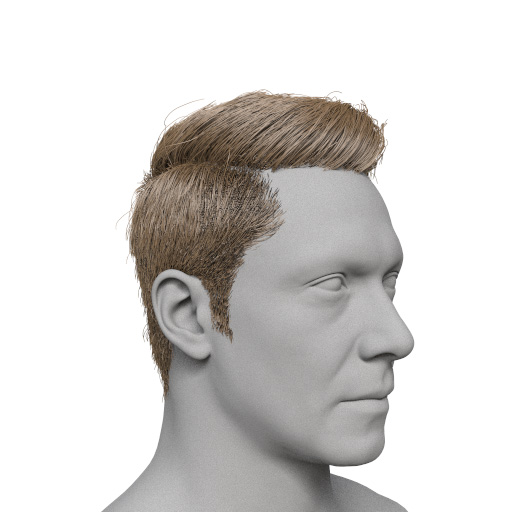} \hfill
\includegraphics[trim =20 0 30 20, clip, width=0.13\linewidth]{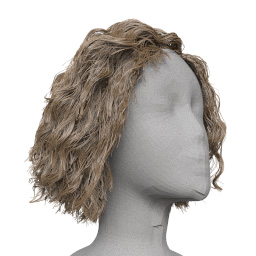} \hfill
\includegraphics[trim =30 50 30 20, clip, width=0.16\linewidth]{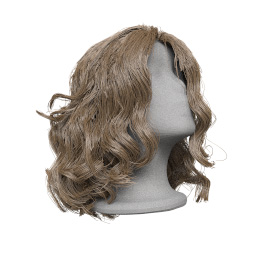} \hfill
\includegraphics[trim =25 0 30 20, clip, width=0.13\linewidth]{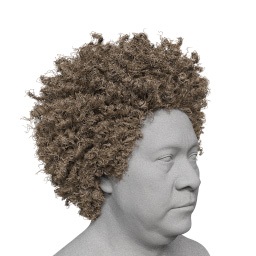} \hfill
\includegraphics[trim =20 0 40 20, clip, width=0.13\linewidth]{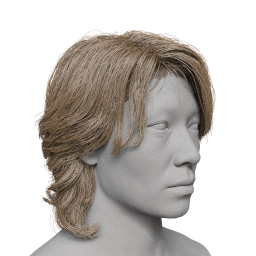} \hfill
\includegraphics[trim =30 0 30 20, clip, width=0.13\linewidth]{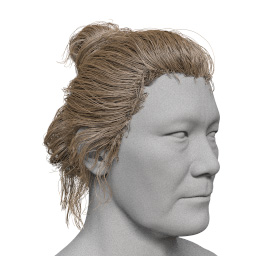} \hfill
\includegraphics[trim =20 20 25 20, clip, width=0.145\linewidth]{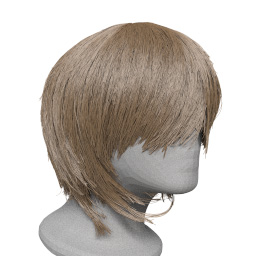}\\
\includegraphics[trim =0 0 80 80, clip, width=0.15\linewidth]{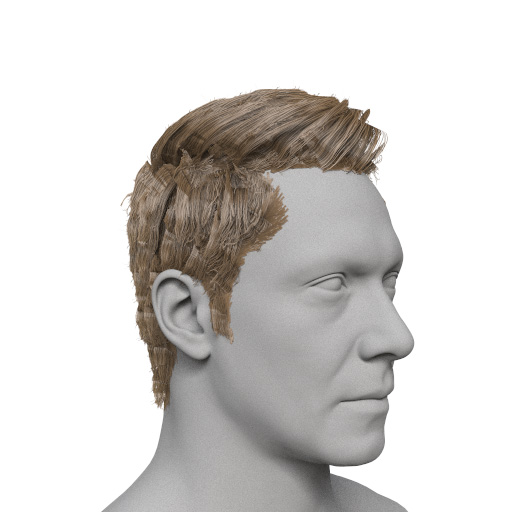} \hfill
\includegraphics[trim =20 0 30 10, clip, width=0.13\linewidth]{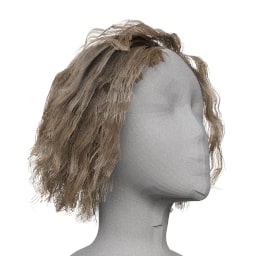} \hfill
\includegraphics[trim =30 50 30 10, clip, width=0.16\linewidth]{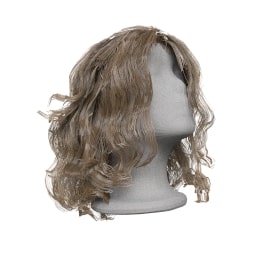} \hfill
\includegraphics[trim =25 0 30 10, clip, width=0.13\linewidth]{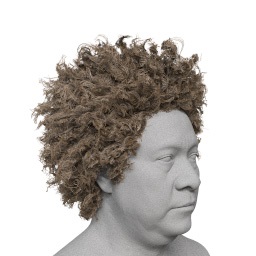} \hfill
\includegraphics[trim =20 0 40 10, clip, width=0.13\linewidth]{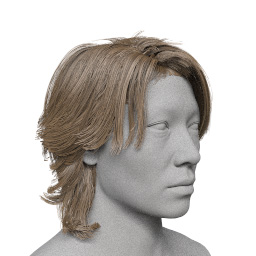} \hfill
\includegraphics[trim =30 0 30 10, clip, width=0.13\linewidth]{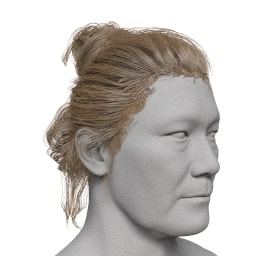} \hfill
\includegraphics[trim =20 20 25 10, clip, width=0.145\linewidth]{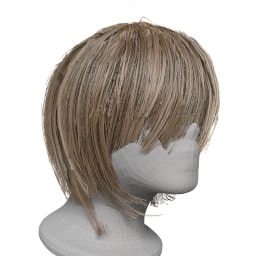} \\
\includegraphics[trim =0 0 40 25, clip, width=0.15\linewidth]{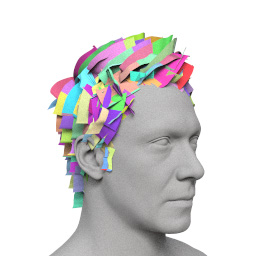} \hfill
\includegraphics[trim =20 0 30 0, clip, width=0.13\linewidth]{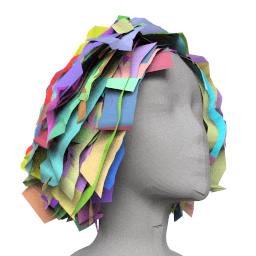} \hfill
\includegraphics[trim =30 50 30 0, clip, width=0.16\linewidth]{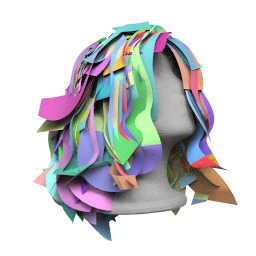} \hfill
\includegraphics[trim =25 0 40 0, clip, width=0.13\linewidth]{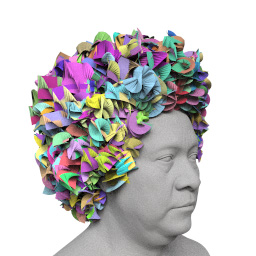} \hfill
\includegraphics[trim =20 0 40 0, clip, width=0.13\linewidth]{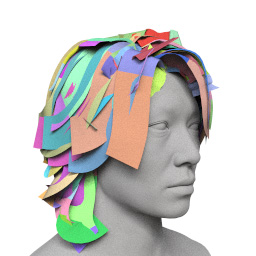} \hfill
\includegraphics[trim =30 0 30 0, clip, width=0.13\linewidth]{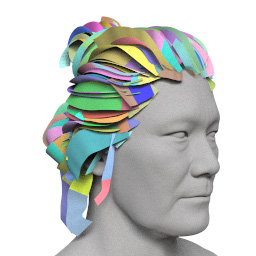} \hfill
\includegraphics[trim =20 20 25 0, clip, width=0.145\linewidth]{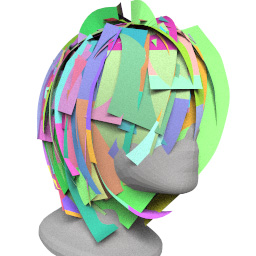} \\
\vspace{-27em}
\begin{flushleft}{
\rotatebox{90}{Our cards only \hspace{2em}  Our cards + textures \hspace{2em} Strands model}\hfill%
}\end{flushleft}
\vspace{1.em}
\figcap{Card \# \quad 200 \quad} \hfill
\figcap{200} \hfill
\figcap{200} \hfill
\figcap{800} \hfill
\figcap{100} \hfill
\figcap{351} \hfill
\figcap{100} \\
\caption{Our automatic pipeline can convert a wide variety of strand-based hairstyles with 40K strands into hair card models using a small number of cards (bottom, depending on hairstyles) and 32 individual card textures, while preserving high visual fidelity. Here, we show the input strand model, our generated cards only, and cards rendered with textures.}
\Description{}
\label{fig:teaser}
\end{teaserfigure}
\maketitle
\input{intro.tex}
\input{related.tex}

\input{problem.tex}
\input{method.tex}
\input{extension.tex}
\input{implementation.tex}
\input{result.tex}

\input{conclusion.tex}

\bibliographystyle{ACM-Reference-Format}
\bibliography{bibliography}


\end{document}

%% file: intro.tex
\section{Introduction}
Hair plays a significant role in creating believable characters and immersive environments in films, video games, and virtual reality. However, the average human scalp has approximately 100,000 to 150,000 hair follicles. Although strand-based hair simulation and rendering have become increasingly popular in recent production pipelines due to their ability to realistically capture dynamic behavior and visual richness~\cite{tafuri2019strand, unrealengine, Huang2023, Hsu2023, Hsu2024, Hsu2025}, hair cards remain a widely adopted technique in real-time applications~\cite{Yibing2016}. Hair cards are flat, textured quad strips that approximate the visual appearance of hair clusters. As illustrated in~\autoref{fig:teaser}, hair cards can effectively reproduce the intricate structure of complex hair styles. Beyond their role in representing high-fidelity hair with a large number of textured planes~\cite{Yibing2016}, hair card representations are also widely used in level-of-detail (LoD), where strand-based models serve as a high-resolution representation, while hair cards can be used as a low-resolution approximation to significantly reduce rendering costs when hair is viewed from a distance. Additionally, mobile games and other low-performance platforms still rely heavily on card-based representations due to strict hardware constraints. Even on high-performance PC platforms, strand-based hair is typically reserved for main characters, while NPCs and background characters use hair cards to reduce runtime costs.

Unfortunately, the industry currently heavily relies on crafting hairstyles manually using hair cards. In this workflow, artists first create a set of hair card textures that contain varying numbers of strands and degrees of curvature. They then manually extrude each hair card from the scalp outward, following the hair flow while varying its length, width, and orientation to achieve a realistic appearance. Apparently, the fidelity quality of the hair card model is primarily bound by the number of cards and textures allowed. As a result, manually crafting a low-resolution hair card model, with a limited number of quads and card textures while maintaining high visual fidelity to the reference strand-based model, is a challenging and labor-intensive task, typically requiring several days to weeks of work from an experienced artist. 

Most recent research works have focused on reconstructing strand-based hair models from single-view images~\cite{zhou2018hairnet, zhang2019hair, wu2022neuralhdhair, zheng2023hairstep}, multi-view inputs~\cite{kuang2022deepmvshair, wu2024monohair, zhou2024groomcap}, and CT scans~\cite{Shen2023CT2Hair}. However, there is a lack of work on converting these reconstructed strands into hair card representations suitable for real-time applications.
Extracting hair cards from strand-based hair models presents several challenges. First, there is a strict budget on both the number of hair cards and on the size of texture, imposed by memory and computational efficiency constraints. Therefore, the extraction process must carefully balance visual fidelity with performance. Second, the hairstyle is highly irregular and chaotic, exhibiting significant variation in strand length, shape, and structure, making it challenging to generate a simplified representation while preserving visual details.
Recently, Unreal Engine (UE)~\cite{unrealengine} introduced an automatic hair card generation tool from strand-based models~\cite{HairCardGen}. However, the quality of its generated results remains insufficient for production use, often failing to preserve the appearance of the target hair strand model when the given hair card target is limited.

This paper presents the first automated pipeline that converts straight and wavy strand-based hair models into hair cards. The process begins with hair strand clustering, where strands are grouped based on a hair shape similarity metric. For each cluster, we optimize the orientation of the card geometry to minimize the strand projection error.
To improve memory efficiency and runtime performance, we perform second-stage clustering on the hair textures, enabling texture sharing across multiple cards. 
Finally, we jointly optimize the hair card positions, strand shapes, and hair widths to minimize differences in tangent, depth, and coverage between our output cards and the original strand-based model.
To this end, conventional RGB-image texture representations can lead to significant aliasing artifacts. Instead, we propose an explicit hair card representation in which each strand is projected into texture space as a 2D curve. These 2D curves serve as an intermediate representation that enables high-quality rendering and supports differentiable optimization. To enhance the visual fidelity for short and coily hairstyles, which is challenging for original hair card representation, our framework also supports baking a hair cap texture and generating a pair of crossed cards per hair cluster, enriching the volumetric appearance of the hair.  

We evaluate our pipeline on a diverse range of hairstyles, including straight, wavy, curly, and coily hair, with a limited number of cards and textures. Experiments show that our automated pipeline surpasses both UE automatic solution and manual-crafted cards.


%% file: related.tex
\section{RELATED WORK}
This section briefly reviews work on hair representation, hair modeling, and extracting planar representations from meshes.

\paragraph{Hair Representation}
Both hair simulation and rendering are computationally expensive due to the large number of individual strands and their complex interactions. To reduce computational costs, researchers have developed various reduced representations over the years, including 2D strips (commonly referred to as \emph{hair cards})~\cite{Koh2001, ward2003modeling}, cubic lattice structures\cite{Volino2006hair}, short hair strips~\cite{Guang2002hair}, and volumetric representations~\cite{wu2016hairmesh, Lee2019volumehair}. During rendering, these reduced representations are expanded into full hair using baked textures or procedural functions. Hair cards remain widely used in the gaming industry due to their simplicity and efficiency~\cite{Yibing2016}. For a comprehensive overview of hair rendering and simulation, we refer readers to the course by~\citet{Bertails2008}. However, creating hair cards for multiple LoDs continues to be a labor-intensive process for artists, representing a significant bottleneck in production pipelines. Recently, \citet{huang2024real} presents a real-time framework to create and render hair LoD dynamically based on pre-clustered hairs.

\begin{figure*}[ht]
\centering
\newcommand{\figcap}[1]{\begin{minipage}{0.163\linewidth}\centering#1\end{minipage}}
\centering
\includegraphics[trim = 20 0 30 0, clip, width=0.16\linewidth]{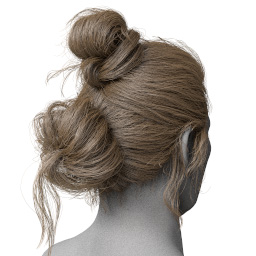} \hfill
\includegraphics[trim = 20 0 30 0, clip, width=0.16\linewidth]{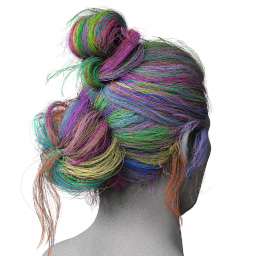} \hfill
\includegraphics[trim = 20 0 30 0, clip, width=0.16\linewidth]{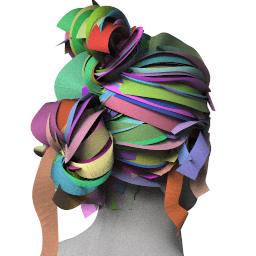} \hfill
\includegraphics[trim = 20 0 30 0, clip, width=0.16\linewidth]{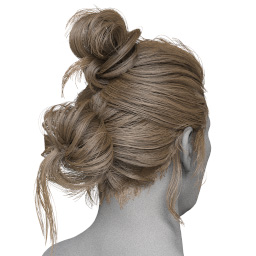} \hfill
\includegraphics[trim = 20 0 30 0, clip, width=0.16\linewidth]{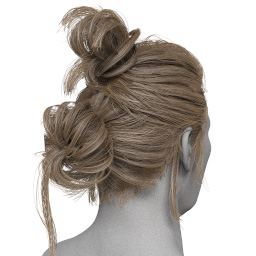} \hfill
\includegraphics[trim = 20 0 30 0, clip, width=0.16\linewidth]{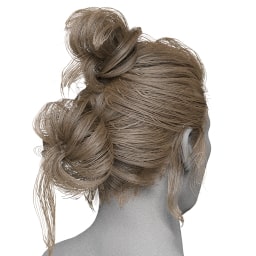}\\ 
\figcap{\small (a) Strands} \hfill
\figcap{\small (b) Clustered strands} \hfill
\figcap{\small (c) Optimized cards} \hfill
\figcap{\small (d) Cards w/ exp. textures} \hfill
\figcap{\small (e) Texture reduction} \hfill
\figcap{\small (f) Final cards w/ textures} \\
\figcap{\small ~} \hfill
\figcap{\small ~} \hfill
\figcap{\small ~} \hfill
\figcap{\small \textcolor{mypsnr}{19.28}/\textcolor{mylpips}{0.148}/\textcolor{mycoverage}{0.026}} \hfill
\figcap{\small \textcolor{mypsnr}{18.54}/\textcolor{mylpips}{0.161}/\textcolor{mycoverage}{0.032}} \hfill
\figcap{\small \textbf{\textcolor{mypsnr}{19.99}/\textcolor{mylpips}{0.140}/\textcolor{mycoverage}{0.021}}}
\caption{Our pipeline: Given the strand-based hair model (a), we first cluster the strands based on their similarity (b). Then, we optimize the orientation for each card (c) and create the corresponding card texture with explicit hair strand representation (d). After hair texture reduction (e), we jointly optimize cards and strands to fine-tune the final result (f). $\textcolor{mypsnr}{\bullet}/\textcolor{mylpips}{\bullet}/\textcolor{mycoverage}{\bullet}$ indicate averaged \textcolor{mypsnr}{PSNR $\uparrow$}, \textcolor{mylpips}{LPIPS $\downarrow$}, and \textcolor{mycoverage}{coverage error $\downarrow$}, respectively. }
\Description{}
\label{fig:pipeline}
\end{figure*}

\paragraph{Hair Modeling}
Traditionally, artists have used tools like Maya XGen to create strand-based and card-based hair models. However, manually crafting hair remains a labor-intensive process. To simplify hair authoring, \citet{Yuksel2009} introduced \emph{hair mesh}, a volumetric representation that provides high-level editing tools for artists. Sketch-based input methods~\cite{fu2007sketching, shen2020deepsketchhair} have also offered user-friendly interaction capability, allowing artists to design hair directly on the screen. For automatic hair modeling, researchers have explored extracting strand-based geometries directly from images. Earlier methods relied on heuristic-based approaches~\rev{\cite{Kong1998generation, paris2008hair, jakob2009capturing, sun2021human,  hu2017avatar}} or leveraged large 3D hair databases~\cite{hu2015single, chai2016autohair, liang2018video} for guidance. Recently, learning-based techniques have demonstrated accuracy and robustness in reconstructing simple hairstyles from single-view images~\cite{zhou2018hairnet, zhang2019hair, wu2022neuralhdhair, zheng2023hairstep} and multi-view inputs~\rev{\cite{kuang2022deepmvshair, rosu2022neural, sklyarova2023neural, wu2024monohair, takimoto2024dr, zakharov2024human, zhou2024groomcap}}.  \citet{Wu2024CurlyCue} present a geometric method to generate highly coiled hair. Beyond explicit modeling and reconstruction, hair synthesis methods enable hairstyle transfer from one to another using feature maps~\cite{Wang09}, further refined with learning-based features~\cite{zhou2023groomgen, sklyarova2024text, chen2024doubly, he2025perm}. However, these techniques primarily aim to generate explicit strand geometries, which are computationally intensive and unsuitable for real-time applications. 
There are also learning-based methods using neural representations \cite{luo2024gaussianhair, zheng2025groomlight, wang2023neuwigs}. While neural-based methods have shown significant progress in reconstructing and representing complex hair geometry, integrating these representations into industrial game engines in a highly efficient manner remains a major challenge.

\paragraph{Billboard Extraction}
In addition to hair, arbitrarily oriented billboards (or impostors) with textures are widely used for rendering trees and forests~\cite{Behrendt2005} and large-scale scenes~\cite{Lall2018, Hladky2022}. To convert a 3D mesh to billboards, \citet{Decoret2003} proposed a greedy optimization algorithm that approximates the 3D model using a discretized plane parameterization in spherical coordinates. This method was later enhanced by \citet{Andujar2004} to produce better-fitting billboards. \citet{Silvennoinen2024} uses a similar concept to generate sets of planars as occluders. However, as hair presents unique challenges due to its extensive strand number and complex geometry, existing billboard generation methods cannot be applied directly.

%% file: problem.tex
\section{Problem Statement and Overview}

In this section, we first provide the problem statement, followed by an overview of our pipeline.

\paragraph{Problem Statement}
The input to our method is a strand-based hairstyle model $\mathcal{H} = \{\mathcal{S}_i\}$, where each strand $\mathcal{S}_i = \{\ss_{i,j}\}$ is represented as a piecewise linear curve consisting of $n^s$ uniformly distributed samples $\ss_{i,j}$ (with the same distance between two consecutive samples on each strand), along with a corresponding head mesh $\mathcal{M}^\text{head}$.
The output of our pipeline is a hair card model $\mathcal{M}^\text{output} = \langle\{\mathcal{C}_i\}_{i=1}^{n^c}, \{\mathcal{T}_i\}_{i=1}^{n^t}\rangle$, composed of $n^c$ hair cards and $n^t$ textures to balance visual fidelity and performance constraints. In order to save texture space, users oftentimes choose $n^c>\!\!>n^t$ such that multiple hair cards need to share the same texture. A hair card $\mathcal{C}_i$ refers to a quad strip with $n^q$ consecutive quads. Our objective is to ensure that the synthesized hair card model $\mathcal{M}^\text{output}$ approximates the visual appearance of the input $\mathcal{H}$ as closely as possible. 

\paragraph{Overview} 
As illustrated in~\autoref{fig:pipeline}, our pipeline begins with hair strand clustering (\autoref{sec:cluster}), where strands are grouped into $n^c$ clusters based on a hair shape similarity metric. For each cluster, we then optimize the card orientation to minimize visual error, producing the initial card geometry (\autoref{sec:cardinit}). Next, we introduce an explicit card texture representation by projecting the 3D curves onto a 2D texture space (\autoref{sec:exptex}). The generated hair textures are further clustered into $n^t$ groups to facilitate texture sharing across cards (\autoref{sec:texReduct}). Finally, we jointly optimize the geometry and strands of all hair cards to minimize the difference between $\mathcal{H}$ and $\mathcal{M}^\text{output}$ using differentiable rendering (\autoref{sec:jointopt}). The optimized cards and strands are then used to generate the final textures via rasterization.

%% file: method.tex
\section{Method}
We revise the steps of our pipeline in this section.

\subsection{\label{sec:cluster}Strand Clustering}
Our method begins by clustering the input strands $\mathcal{H}$ into a number of subsets so each cluster can be represented as a hair card. We use a clustering method proposed by~\cite{Wang09}. Specifically, a hair shape similarity metric is defined as $\gamma(\mathcal{S}_i, \mathcal{S}_j) = \sum_{k=1}^{n^s} \|\ss_{i,k} - \ss_{j,k}\|_2^2 / n^s$ to quantify the sample-wise distance between two hair strands $\mathcal{S}_i$ and $\mathcal{S}_j$. Using this metric, we group $\mathcal{H}$ into clusters $\{\mathcal{G}_k\}$ with a k-mean clustering. Clearly, the center of each cluster is a mean strand denoted as $\bar{\mathcal{S}}_k=\{\bar{\ss}_{k,j}\}$ with samples being $\bar{\ss}_{k,j}=\sum_{\mathcal{S}_i\in\mathcal{G}_k}\ss_{i,j}/|\mathcal{G}_k|$, where $\mathcal{G}_k\subseteq\mathcal{H}$ is a strand subset.

\subsection{Card Geometry Initialization} \label{sec:cardinit} 
Given the strand clusters ${\mathcal{G}_k}$, our next step is to initialize the hair card and prepare for further optimization. We first initialize an orthogonal frame along the mean strand and build initial card geometry. Then, we optimize orientation for each card to minimize projection error.

\setlength{\columnsep}{1.5em}
\begin{wrapfigure}{r}{0.41\linewidth}
\centering
\vspace{-1.5em}
\includegraphics[width=\linewidth, trim =50 30 30 80, clip]{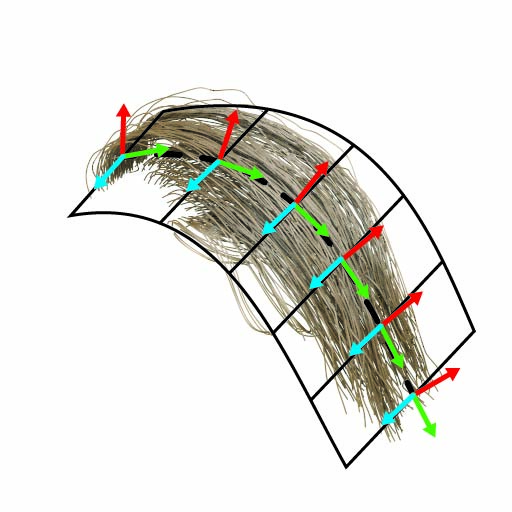} 
\vspace{-1.5em}
\caption{Example of hair cluster and its hair strip as well as per-sample frames.}
\label{fig:frame}
\vspace{-1.5em}
\end{wrapfigure}
\paragraph{Card Geometry Fitting}
To initialize the card geometry, we need to build an orthogonal frame system $\{\tt_{k,j}, \nn_{k,j}, \bb_{k,j}\}$ at each sample $\bar\ss_{k,j}$ along the central strand $\bar{\mathcal{S}}_k$, where $\tt_{k,j}, \nn_{k,j}, \bb_{k,j}$ are the tangent, normal, and binormal vectors, respectively. The problem of building a frame system for a 3D curve has been thoroughly studied, for which the standard construction is the Frenet-Serret formulas. 
But this formula suffers from singular configurations, and a more numerically stable choice is the Bishop formulas~\cite{Bergou2008DER}, which have been adopted in strand-based hair simulation. Specifically, given the normal $\nn_{k,1}$ for the first curve segment, the entire frame system can be computed by solving the Bishop formulas, and we refer readers to~\cite{Bergou2008DER} for the piecewise linear discretization.
Given the frame system, we calculate the maximum distance from $\bar\ss_{k,j}$ to all corresponding strand samples within the cluster along the binormal direction such that: 
\begin{align}
W_{k,j} = \max_{\mathcal{S}_i \in \mathcal{G}_k} \left| (\ss_{i,j} - \bar\ss_{k,j}) \cdot \bb_{k,j} \right|. 
\end{align}
Then, the card is constructed by connecting all $\pp_{k,j}^\pm=\bar{\ss}_{k,j}\pm W_{k,j} \bb_{k,j}$ along the positive and negative $\bb_{k,j}$ at each $\bar\ss_{k,j}$ to form a quad strip with $n^s$ quads. Finally, we downsample the quad strip to use only $n^q$ quads, leading to our card geometry $\mathcal{C}_k$ as shown in~\autoref{fig:frame}.

\paragraph{Card Orientation Optimization}
In the above discussion, we have assumed that the root normal $\nn_{k,1}$ is given as the boundary condition of the Bishop formulas. This root normal $\nn_{k,1}$ for the first card segment needs to be carefully computed to provide a good initial guess for further optimizations. We propose to optimize $\nn_{k,1}$ by minimizing the difference between the strand cluster $\mathcal{G}_k$ and the hair card $\mathcal{C}_k$. To this end, we introduce the projection operator $\ss_{i,j}^\Gamma=\text{argmin}_{\ss\in\mathcal{C}_k}\|\ss-\ss_{i,j}\|$ as finding the closest point $\ss$ on the quad strip of hair card $\mathcal{C}_k$, that is closest to $\ss_{i,j}$. We then define the projection error as follows:
\begin{align}
\label{eq:cardopt}
L^\text{proj} = \sum_{\mathcal{S}_i\in\mathcal{G}_k}\sum_{\ss_{i,j}\in\mathcal{S}_i}\|\ss_{i,j}^\Gamma-\ss_{i,j}\|.
\end{align}
Although the Bishop formulas are differentiable, the projection operator is non-differentiable, so gradient-based continuous optimization is non-available to minimize $L^\text{proj}$. Fortunately, the solution space is rather small. Since~\autoref{eq:cardopt} is independent for different cards and due to the condition that $\nn_{i,1}$ is orthogonal to $\tt_{i,1}$, $\nn_{i,1}$ essentially lies on a 3D circle. Therefore, we evenly sample a fixed number of potential $\nn_{i,1}$ along the 3D circle and pick the direction leading to the smallest value of~\autoref{eq:cardopt}. At this point, we have initialized all the card geometries.

\subsection{Explicit Hair Card Texture} \label{sec:exptex}
In the standard rendering pipeline, textures are introduced for each hair card to represent various attributes in the form of RGB images. However, the hair strands are extremely thin, and rasterizing them into RGB textures would introduce severe aliasing error and noisy gradient information. To mitigate this issue, we propose a novel intermediary explicit hair card representation.
Specifically, each hair strand $\mathcal{S}_i\in\mathcal{G}_k$ is first projected onto the quad strip $\mathcal{C}_k$. Specifically, for each $\ss_{i,j}\in\mathcal{S}_i$, we find the closest point $\ss_{i,j}^\Gamma$ lying on the associated quad strips $\mathcal{C}_k$. The corresponding 2D uv-coordinate $\ss_{i,j}^\text{uv} = (u_{i,j}, v_{i,j}) \in [0,1]^2$ on the texture can be computed such that the 3D world position of each sample $\ss_{i,j}$ can be reconstructed as:
\begin{align} 
\label{eq:uv2world}
\ss_{i,j} = u_{i,j} \pp_{i,j}^0 + v_{i,j} \pp_{i,j}^1 + (1 - u_{i,j} - v_{i,j}) \pp_{i,j}^2 + z \nn_{i,j}^\star,
\end{align}
where $\pp_{i,j}^\bullet$ are the vertices of the corresponding triangle face in the card geometry that contains $\ss_{i,j}^\Gamma$, $\nn_{i,j}^\star$ is the normal of that triangle, and $z\in\mathbb{R}$ is the displacement along $\nn_{i,j}^\star$. Hence, each hair card texture can be explicitly represented as a set of points $\{\ss_{i,j}^\text{uv}\}$ embedded within the uv-space.  Using $\{\ss_{i,j}^\text{uv}\}$ as our intermediary representation induces two remarkable features. First, this representation is amenable to end-to-end differentiable rendering, while it does not suffer from aliasing error induced by RGB images. Indeed, for differentiable rendering, given the card mesh and explicit hair card texture, 3D hair strands can be first recovered from~\autoref{eq:uv2world}. The user-defined per-strand hair width is denoted as $w_i$. Each and every step of this procedure is differentiable, and a similar technique is proposed in~\cite{sklyarova2023neural}. Further, our representation can be converted back to RGB images via rasterization. Finally, to avoid uv-space tangent distortion by the card geometry in world space, we decompose the tangent from strand geometry and store it as a per-vertex 3D attribute denoted as $\{\tt_{i,j}^\text{3D}\}$, which plays a central role in the hair appearance model. The renderer can also produce depth and coverage maps by utilizing depth and flat color as attributes for each vertex.

\subsection{Texture Reduction} \label{sec:texReduct}
To facilitate texture sharing and improve both memory efficiency and runtime performance, users could optionally reduce the number of textures before packing them into a texture atlas. Clusters with similar strand counts and shapes are grouped to share the same texture. Specifically, after the per-cluster optimization, we allow users to provide the target texture number $n^t$. We then perform another round of k-means clustering to merge cards with similar textures. To this end, we first rasterize the UV-space strands into RGB textures and then compute the Learned Perceptual Image Patch Similarity (LPIPS) metric~\cite{Zhang2018} between all pairs of card textures. The LPIPS metric is used to guide our k-means clustering. For each cluster, only the texture closest to the k-mean center is retained, and it is reused across all hair cards within the same cluster. After the clustering, our joint optimization would then fine-tune the shared textures to collectively match the appearance. 

\subsection{Joint Optimization} \label{sec:jointopt}
After texture reduction, we perform another optimization with the visual losses and geometric regularization to match the collective visual appearance between the input strands $\mathcal{H} = \{\mathcal{S}_i\}$ and our final cards $\{\mathcal{C}_i\}$ with uv-space strands $\{\ss_{i,j}^\text{uv}\}$ and tangent $\{\tt_{i,j}^\text{3D}\}$.

\paragraph{Optimization}
Given the differentiable renderer, we fine-tune both the card geometry and strand textures for each cluster using gradient-based optimization by solving the following optimization:
\begin{align} 
\label{eq:totalloss}
\underset{\pp_{k,j}^\pm,~\ss_{i,j}^\text{uv},~\tt_{i,j}^\text{3D},~w_{i}}{\text{argmin}} \quad \sum_v \lambda^v L^v + \sum_g \lambda^g L^g ,
\end{align}
where quad strip position $\pp_{k,j}^\pm$, strand texture coordinates $\ss_{i,j}^\text{uv}$, tangent $\tt^\text{3D}_{i,j}$, and per-strand width $w_i$ are all included as decision variables. To ensure high output quality, we combine visual loss terms $v \in \{\text{tangent},~\text{depth},~\text{dice}\}$ and geometry regularization terms $g \in$ $\{\text{match},~\text{collision}\}$, which are detailed below.

\paragraph{Visual Losses} 
While our ultimate goal is to produce a hair card model that looks identical to the input strand model, rendering complicated light interactions between a large number of hair segments is computationally intractable. In particular, the interaction between light and a hair strand is described by the Bidirectional Curve Scattering Distribution Function (BCSDF) denoted $f(\theta_i, \phi_i, \theta_o, \phi_o)$, as proposed by~\citet{marschner2003light}. Here, $\langle \theta_i,\phi_i \rangle$ and $\langle \theta_o,\phi_o \rangle$ represent the incoming and outgoing light directions, respectively. The inclination angle $\theta_\bullet$ represents the deviation from the strand normal plane, while the azimuth angle $\phi_\bullet$ captures the orientation around the hair axis, both derived from the strand tangent. In order for more efficient optimization, instead of matching the final appearance, our optimization matches the tangent channel for shading and the depth channel for strand position along the view direction. In particular, tangent and depth losses are measured by the Mean-Squared Error ($\mathsf{MSE}$) over all views as:
\begin{align}
L^{v} = \int_{\mathbb{S}^2} \mathsf{MSE}\bigl(\mathcal{I}^{v}(\mathcal{H},\omega), \mathcal{I}^{v}(\langle\mathcal{C}_k, \ss_{i,j}^\text{uv},\tt_{i,j}^\text{3D},w_i\rangle,\omega)\bigr) d \omega,
\end{align}
where $\mathcal{I}^v(\bullet,\omega)$ is the rendering function that rasterizes the channel ${v \in \{\text{tangent},~\text{depth}\}}$ of entities $\bullet$ under view direction $\omega$. We further add the following dice loss~\cite{Sudre2017} for matching the silhouette:
\begin{align}
L^\text{dice}=\int_{\mathbb{S}^2}1 - D\bigl(\mathcal{I}^\text{mask}(\mathcal{H},\omega), \mathcal{I}^\text{mask}(\langle\mathcal{C}_k,\ss_{i,j}^\text{uv},w_i\rangle,\omega)\bigr) d \omega,
\end{align}
where $D(A, B) = {2 |A \cap B |}/{(|A| + |B|)}$ is the Sørensen-Dice coefficient to quantify the similarity between two masks, $A$ and $B$, where  $|A|$ and $|B|$ are the total number of pixels in $A$ and $B$, and $|A \cap B |$ is the common area between $A$ and $B$. For rendering the binary mask image, we set all pixels to one in a zero background. Note that we do not need the tangents $\tt_{i,j}^\text{3D}$ for mask rendering.

\paragraph{Geometric Regularization}
For accurate visual appearance, we treat $\tt^\text{3D}_{i,j}$ as a separate decision variable that is not bound to the strand geometry. However, we still encourage the geometric definition of tangent is consistent with the optimized tangent direction, via the regularization:
\begin{align}
L^\text{match} = \sum_{\mathcal{G}_k}\sum_{\mathcal{S}_i\in\mathcal{G}_k}\sum_{j=1}^{N^s}\left\| \tt^\text{3D}_{i,j} - \frac{\ss_{i,j+1} - \ss_{i,j}}{\|\ss_{i,j+1} - \ss_{i,j}\|}\right\|^2.
\end{align}
Finally, we introduce a collision loss to prevent hair cards from penetrating the head mesh, formulated as:
\begin{align}
L^\text{collision} = \sum_{\mathcal{G}_k}\sum_{j=1}^{N^s}\sum_{\bullet\in\{+,-\}}
\|\min(0,\text{SDF}(\pp_{k,j}^\bullet, \mathcal{M}^\text{head}))\|^2,
\end{align}
where $\text{SDF}$ is the signed distance from a point to the head mesh.

\begin{figure}[ht!]
\newcommand{\figcap}[1]{\begin{minipage}{0.24\linewidth}\centering#1\end{minipage}}
\centering
\includegraphics[trim =0 200 0 0, clip, width=0.243\linewidth]{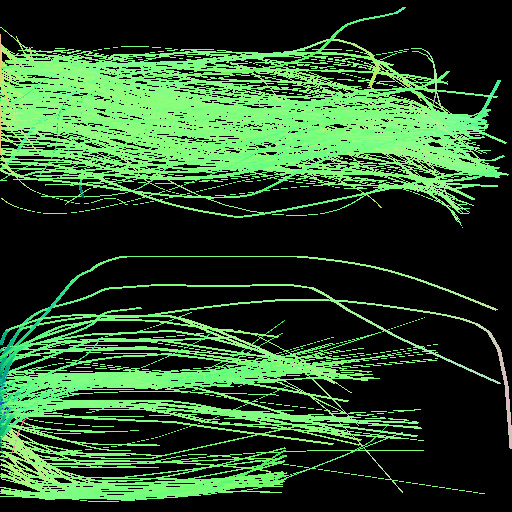} \hfill
\includegraphics[trim =0 200 0 0, clip, width=0.243\linewidth]{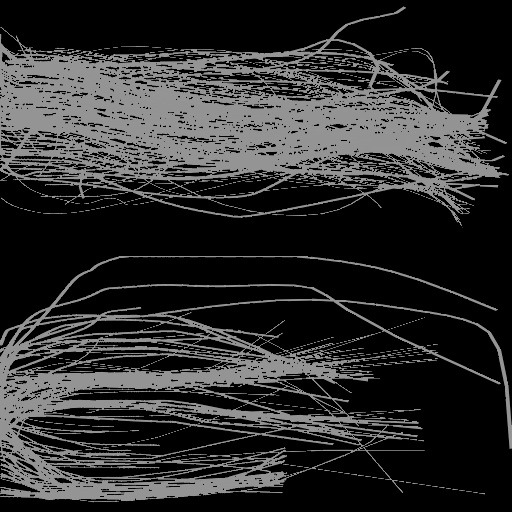} \hfill
\includegraphics[trim =0 200 0 0, clip, width=0.243\linewidth]{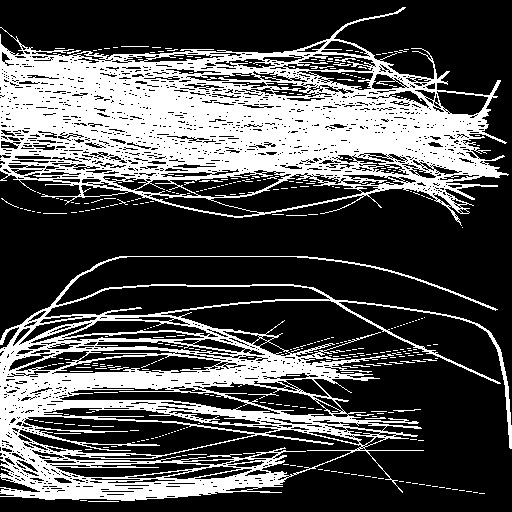} \hfill
\includegraphics[trim =0 200 0 0, clip, width=0.243\linewidth]{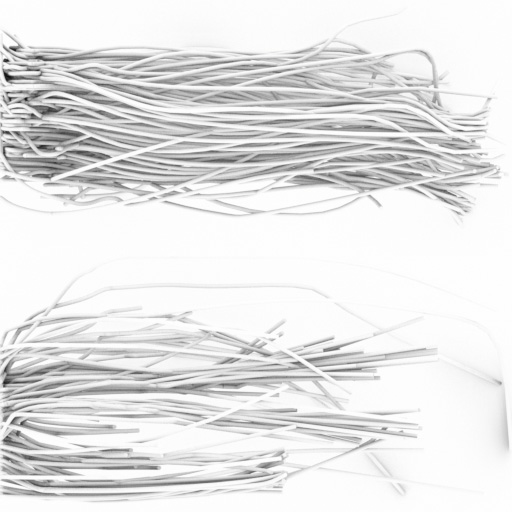}\\ 
\figcap{\small Tangent } \hfill
\figcap{\small Depth} \hfill
\figcap{\small Alpha} \hfill
\figcap{\small AO}
\caption{\label{fig:texture} Example of our output hair card textures }
\Description{}
\end{figure}

\paragraph{Texture Baking}
As our final step, after optimization, we rasterize the UV-space strands and tangents into RGB-format textures to generate the final tangent, depth, and alpha maps, as shown in~\autoref{fig:texture}. Additionally, we bake an ambient occlusion (AO) texture to enhance visual detail and store local lighting information on the hair cards. We convert explicit hair card texture into 3D tubes in UV-space and illuminate them with a directional light along the z-direction. Next, we perform standard offline ray-traced AO baking, computing surface color contributions after multiple light bounces. Finally, the resulting shaded surface is projected back onto the card plane to create the AO texture. Note that while it is possible to also optimize the $z$ offset along the normal direction in~\autoref{eq:uv2world}, we found in practice that keeping it fixed during optimization can provide sufficient results.

%% file: extension.tex
\subsection{Extensions}
To enhance both the visual fidelity of our output and the practical usability of our pipeline, we introduce two additional optional features.

\begin{figure}[ht!]
\newcommand{\figcap}[1]{\begin{minipage}{0.24\linewidth}\centering#1\end{minipage}}
\centering
\includegraphics[trim =110 70 110 60, clip, width=0.24\linewidth]{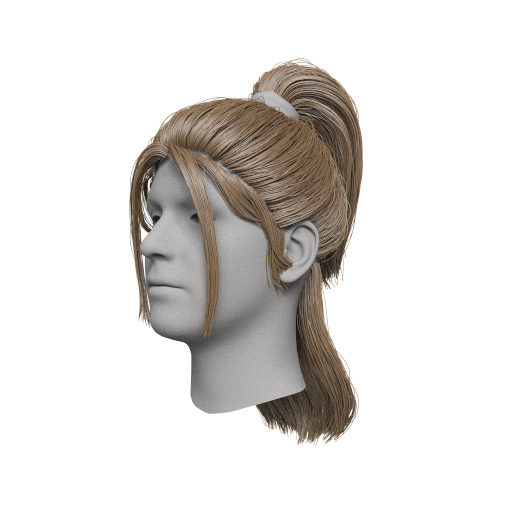} \hfill
\includegraphics[trim =110 70 110 60, clip, width=0.24\linewidth]{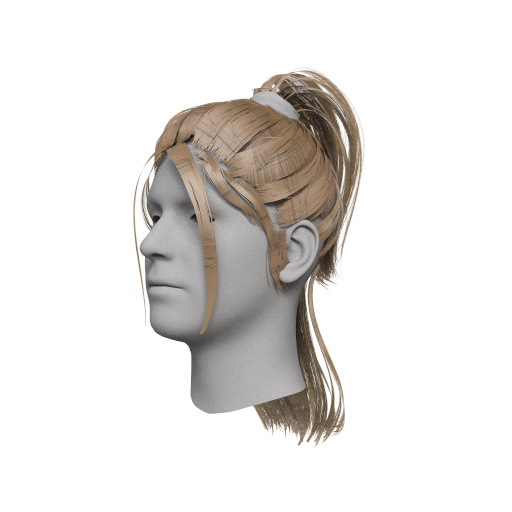} \hfill
\includegraphics[trim =110 70 110 60, clip, width=0.24\linewidth]{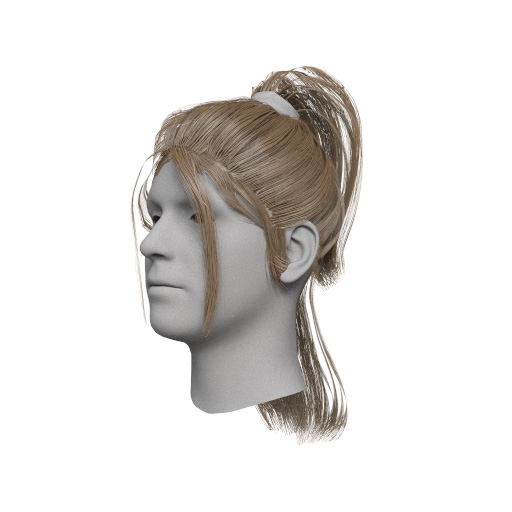} \hfill
\includegraphics[trim =110 70 110 60, clip, width=0.24\linewidth]{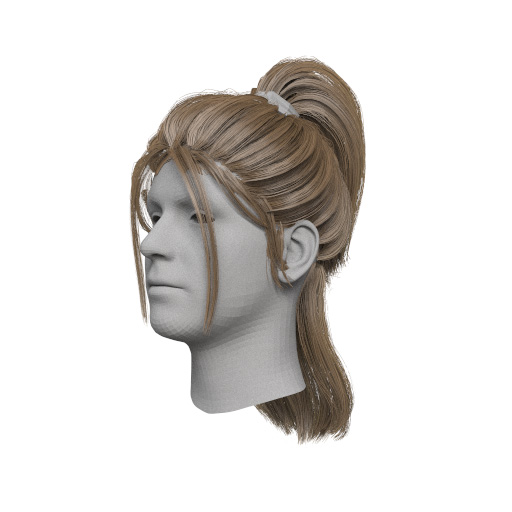}\\ 
\figcap{\small Strands} \hfill
\figcap{\small UE cards} \hfill
\figcap{\small Our cards} \hfill
\figcap{\small Our crossed}
\caption{\label{fig:ponytail} Compared with single cards from Unreal Engine auto tool~\cite{HairCardGen} and ours, our crossed cards can create a volumetric appearance, especially around the ponytail. }
\Description{}
\end{figure}

\paragraph{Crossed Cards}
A common limitation of using billboards/cards is that when the view direction becomes nearly parallel to the card plane, the card appears invisible. To simulate hair volume and improve visual fidelity from multiple viewing angles, we introduce crossed hair cards during generation. Specifically, for each cluster, we generate a pair of hair cards placed at a 90-degree angle to each other, creating the illusion of volumetric hair using flat geometry, as shown in~\autoref{fig:ponytail}. In practice, during the card generation stage (\autoref{sec:cardinit}), we simply create an additional card perpendicular to the primary card at each segment by rotating both $\nn_{k,j}$ and $\tt_{k,j}$ over 90 degrees, giving a crossed configuration.

\begin{figure}[b!]
\centering
\newcommand{\figcap}[1]{\begin{minipage}{0.24\linewidth}\centering#1\end{minipage}}
\includegraphics[trim =80 0 80 75, clip, width=0.24\linewidth]{figs/short/strand.jpg} \hfill
\includegraphics[trim =80 0 80 75, clip, width=0.24\linewidth]{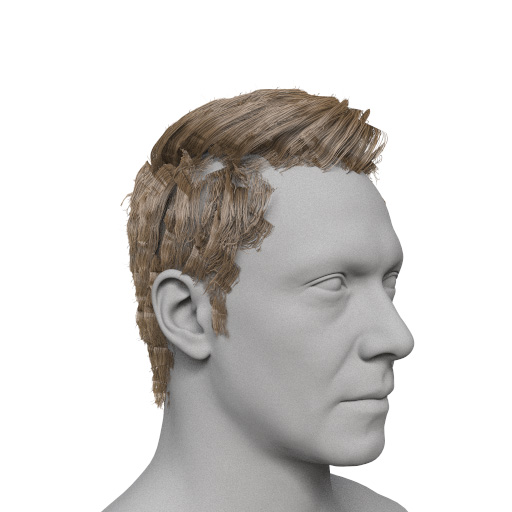} \hfill
\includegraphics[trim =80 0 80 75, clip, width=0.24\linewidth]{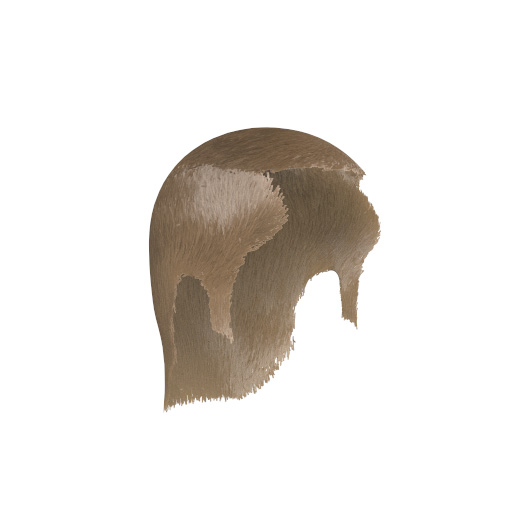} \hfill
\includegraphics[trim =80 0 80 75, clip, width=0.24\linewidth]{figs/short/our_card_cap.jpg}\\ 
\figcap{\small Strands} \hfill
\figcap{\small Our crossed} \hfill
\figcap{\small Our cap} \hfill
\figcap{\small Our crossed + cap} 
\caption{\label{fig:short}Our hair cap can improve the scalp coverage and hairline.} 
\Description{}
\end{figure}

\paragraph{Hair Cap}
Since hair cards alone may not fully cover the scalp under certain camera angles or lighting conditions, particularly near the roots, we construct a hair cap, a base layer of geometry and texture applied to the scalp that simulates the appearance of short, dense hair, so it can efficiently represent root fuzz and base hair density with minimal computational cost.
To create the hair cap mesh and its associated texture, we begin by projecting all hair roots onto their nearest faces on the head mesh $\mathcal{M}^\text{head}$. We then extract all faces containing at least one hair root, along with their one-ring neighboring faces, to form the hair cap mesh. In order to prevent z-fighting during rendering, we slightly extrude the vertices of the cap mesh along their normal directions by a small distance $\epsilon^\text{cap}$. Finally, for each strand, we extract the first sub-segment with a fixed arc-length of $\epsilon^\text{root}$ and bake the sub-segment onto the scalp texture. Specifically, we bake the tangent, alpha, and AO into three texture channels as shown in~\autoref{fig:short}.

%% file: implementation.tex
\section{Implementation Details}
This section provides a detailed description of the implementation of our rendering pipeline employed in both the optimization stage and the final results presentation.

\paragraph{Optimization Stage}
After reconstructing all strand control points using \autoref{eq:uv2world}, the strands are rasterized into camera-oriented quad strips with a user-defined strand width through Nvdiffrast \cite{Laine2020diffrast}. Since our optimization is guided by loss functions defined over tangent, depth, and coverage, no hair BCSDF is employed for shading. Instead, the renderer produces these G-buffer maps with the corresponding attributes for evaluating the loss function.

\paragraph{Rendering Final Results}
To ensure high-quality rendering results, we employ an offline path tracing pipeline for final image generation. Hair appearance is modeled using the classical formulation of hair BCSDF~\cite{marschner2003light}, and the local frames are reconstructed from tangent and depth textures during BCSDF sampling and evaluation. Upon ray–card intersection, the alpha texture is used to determine the strand occupancy at the texel level. Hair-hair occlusion is separated into two stages: inter-card occlusion is directly resolved via ray tracing, while intra-card strand occlusion is approximated using our precomputed AO texture.

%% file: result.tex
\begin{table*}[t!]
\centering
\caption{Statistics about card \# used in both our method and Unreal Engine's auto cards, as well as the corresponding visual metric statistics, \textcolor{mypsnr}{PSNR $\uparrow$}, \textcolor{mylpips}{LPIPS $\downarrow$}, \textcolor{mycoverage}{coverage error $\downarrow$}, and \rev{chamfer distance (CD)} $\downarrow$. Crossed card is indicated by $\times2$.} \label{tab:performance}
\scalebox{1}
{
\setlength{\tabcolsep}{3px}
\begin{tabular}{lcc|cccc|cccc}
\toprule
            &  &   & \multicolumn{4}{c|}{Unreal Engine~\cite{HairCardGen}}     &  \multicolumn{4}{c}{Ours} \\
Model       & Fig.                   \#  & Card \# & PSNR $\uparrow$ & LPIPS $\downarrow$ & Coverage $\downarrow$& \rev{CD} $\downarrow$ & PSNR $\uparrow$ & LPIPS $\downarrow$ & Coverage $\downarrow$ & \rev{CD} $\downarrow$ \\
\midrule
Ponytail    &\ref{fig:ponytail}   &  100$\times$2 & \textcolor{mypsnr}{20.91} & \textcolor{mylpips}{0.123} & \textcolor{mycoverage}{0.022}  & 0.003 & \textbf{\textcolor{mypsnr}{22.70}} & \textbf{\textcolor{mylpips}{0.095}} & \textbf{\textcolor{mycoverage}{0.010}}& 0.003 \\
Short       &\ref{fig:short}      &  100$\times$2 & \textcolor{mypsnr}{22.34} & \textcolor{mylpips}{0.112} & \textcolor{mycoverage}{0.014} & 0.007  & \textbf{\textcolor{mypsnr}{23.51}} & \textbf{\textcolor{mylpips}{0.100}} & \textbf{\textcolor{mycoverage}{0.010}} & 0.004 \\
Straight    &\ref{fig:straight}   &  100 & \textcolor{mypsnr}{17.39} & \textcolor{mylpips}{0.230} & \textcolor{mycoverage}{0.038} & 0.005 & \textbf{\textcolor{mypsnr}{18.21}} & \textbf{\textcolor{mylpips}{0.176}} & \textbf{\textcolor{mycoverage}{0.034}} & 0.005 \\
Curly       &\ref{fig:gallary}        &  400 & \textcolor{mypsnr}{18.54} & \textcolor{mylpips}{0.153} & \textcolor{mycoverage}{0.038} & 0.005  & \textbf{\textcolor{mypsnr}{20.89}} & \textbf{\textcolor{mylpips}{0.138}} & \textbf{\textcolor{mycoverage}{0.016}} & 0.005 \\
Bangs       &\ref{fig:gallary}       &  100 & \textcolor{mypsnr}{18.09} & \textcolor{mylpips}{0.218} & \textcolor{mycoverage}{0.032} & 0.006  & \textbf{\textcolor{mypsnr}{20.47}} & \textbf{\textcolor{mylpips}{0.154}} & \textbf{\textcolor{mycoverage}{0.018}} & 0.006 \\
Blowout     &\ref{fig:gallary}        &  200 & \textcolor{mypsnr}{17.16} & \textcolor{mylpips}{0.200} & \textcolor{mycoverage}{0.038} & 0.005 & \textbf{\textcolor{mypsnr}{20.44}} & \textbf{\textcolor{mylpips}{0.162}} & \textbf{\textcolor{mycoverage}{0.017}} & 0.005 \\
Wavy        &\ref{fig:gallary}       &  200 & \textcolor{mypsnr}{16.82} & \textcolor{mylpips}{0.228} & \textcolor{mycoverage}{0.041} & 0.006  & \textbf{\textcolor{mypsnr}{19.89}} & \textbf{\textcolor{mylpips}{0.166}} & \textbf{\textcolor{mycoverage}{0.016}} & 0.006 \\
Bun         &\ref{fig:bun}        &  351 & \textcolor{mypsnr}{17.78} & \textcolor{mylpips}{0.209} & \textcolor{mycoverage}{0.038} & 0.004 & \textbf{\textcolor{mypsnr}{19.99}} & \textbf{\textcolor{mylpips}{0.140}} & \textbf{\textcolor{mycoverage}{0.021}} & 0.004 \\
Braid       &\ref{fig:braid}      &  400 & \textcolor{mypsnr}{23.25} & \textcolor{mylpips}{0.086} & \textcolor{mycoverage}{0.010} & 0.003 & \textbf{\textcolor{mypsnr}{24.90}} & \textbf{\textcolor{mylpips}{0.077}} & \textbf{\textcolor{mycoverage}{0.007}} & 0.003 \\
Fringe       &\ref{fig:multiview}      &  100 & \textcolor{mypsnr}{18.47} & \textcolor{mylpips}{0.166} & \textcolor{mycoverage}{0.036} & 0.007 & \textbf{\textcolor{mypsnr}{19.24}} & \textbf{\textcolor{mylpips}{0.158}} & \textbf{\textcolor{mycoverage}{0.029}} & 0.008 \\
Coily       &\ref{fig:coily}      &  400$\times$2 & \textcolor{mypsnr}{17.90} & \textcolor{mylpips}{0.174} & \textcolor{mycoverage}{0.032} & 0.007 & \textbf{\textcolor{mypsnr}{19.09}} & \textbf{\textcolor{mylpips}{0.131}} & \textbf{\textcolor{mycoverage}{0.022}} & 0.007 \\
\bottomrule
\end{tabular}
}
\end{table*}

\section{Results}
We conduct all experiments on a system equipped with an AMD Ryzen Threadripper 3970X 32-core CPU, 256 GB of memory, and an NVIDIA RTX 3090 GPU. Our framework is implemented in PyTorch and customizes Nvdiffrast~\cite{Laine2020diffrast} as our backbone differentiable render. 

We evaluate our pipeline on a diverse set of hairstyles, including short, bangs, straight, ponytail, bun, wavy, curly, and coily, from MetaHuman~\cite{unrealengine} and a dataset in previous high-fidelity hair reconstruction work~\cite{Shen2023CT2Hair}. All hair models are down-sampled to use 40K hair strands, each with $n^s = 32$ samples. Following the hair card texture settings used in MetaHuman, we set our output texture atlas size to $2048 \times 2048$, which accommodates $n^t = 4 \times 8 = 32$ individual hair card textures, each at a resolution of $512 \times 256$ by default. 
The card orientation optimization uses a screen size $128 \times 128$, while the joint optimization uses $256 \times 256$ to capture strand-level details. For card geometry fitting, we sample 20 candidate normals evenly distributed along a positive half-circle. During optimization, we uniformly sample 64 viewpoints on the unit sphere, oriented toward the origin, to balance between coverage and computational efficiency. We set $n^q$ to 31 for standard cards and 15 for cross cards, except for Afro, we use 48. 
To assess the visual similarity to the input strand-based model, we use three averaged metrics, as presented in~\autoref{tab:performance}. \textcolor{mypsnr}{PSNR $\uparrow$} for the shading evaluation, \textcolor{mylpips}{LPIPS $\downarrow$} for perceptual similarity, especially high-frequency hair details, and \textcolor{mycoverage}{coverage error $\downarrow$} for the hair silhouette. We also use chamfer distance $\downarrow$ to evaluate the geometric accuracy. The average values of these metrics are computed over 12 uniformly distributed viewpoints around the hair.

\begin{figure}[t!]
\newcommand{\figcap}[1]{\begin{minipage}{0.19\linewidth}\centering#1\end{minipage}}
\centering
\figcap{\small Strands \\ ~} \hfill
\figcap{\small Unreal Engine's cards} \hfill
\figcap{\small Ours 16 textures} \hfill
\figcap{\small Ours 32 textures} \hfill
\figcap{\small Ours 64 textures} \\ 
\includegraphics[trim = 20 0 40 20, clip, width=0.19\linewidth]{figs/taro/strand.jpg} \hfill
\includegraphics[trim = 20 0 40 20, clip, width=0.19\linewidth]{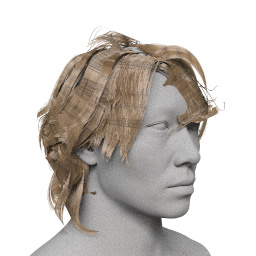} \hfill
\includegraphics[trim = 20 0 40 20, clip, width=0.19\linewidth]{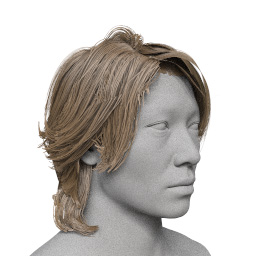} \hfill
\includegraphics[trim = 20 0 40 20, clip, width=0.19\linewidth]{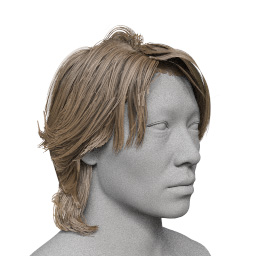}\hfill
\includegraphics[trim = 20 0 40 10, clip, width=0.19\linewidth]{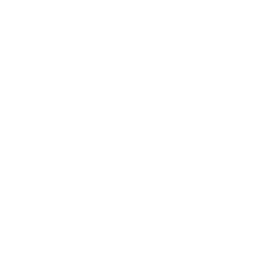} \\
\figcap{\small Reference} \hfill
\figcap{\tiny \textcolor{mycard}{50}/\textcolor{mytex}{32}, \textcolor{mypsnr}{16.97}/\textcolor{mylpips}{0.270}/\textcolor{mycoverage}{0.048} } \hfill
\figcap{\tiny \textcolor{mycard}{50}/\textcolor{mytex}{16}, \textcolor{mypsnr}{17.66}/\textcolor{mylpips}{0.203}/\textcolor{mycoverage}{0.045} } \hfill
\figcap{\tiny \textcolor{mycard}{50}/\textcolor{mytex}{32}, \textbf{\textcolor{mypsnr}{17.99}/\textcolor{mylpips}{0.184}/\textcolor{mycoverage}{0.40}}} \hfill
\figcap{\small \vspace{-8em} N/A ~} \\
\includegraphics[trim = 20 0 40 10, clip, width=0.19\linewidth]{figs/taro/empty.jpg} \hfill
\includegraphics[trim = 20 0 40 10, clip, width=0.19\linewidth]{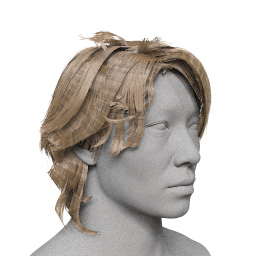} \hfill
\includegraphics[trim = 20 0 40 10, clip, width=0.19\linewidth]{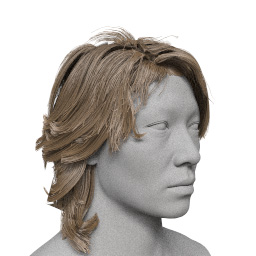} \hfill
\includegraphics[trim = 20 0 40 10, clip, width=0.19\linewidth]{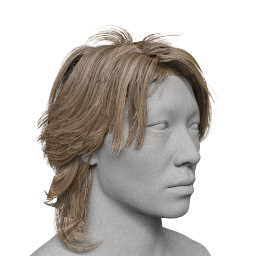} \hfill
\includegraphics[trim = 20 0 40 10, clip, width=0.19\linewidth]{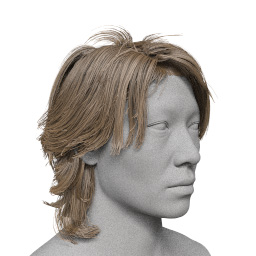}\\ 
\figcap{\tiny ~} \hfill
\figcap{\tiny \textcolor{mycard}{100}/\textcolor{mytex}{64}, \textcolor{mypsnr}{17.60}/\textcolor{mylpips}{0.214}/\textcolor{mycoverage}{0.035}} \hfill
\figcap{\tiny \textcolor{mycard}{100}/\textcolor{mytex}{16}, \textcolor{mypsnr}{18.03}/\textcolor{mylpips}{0.192}/\textcolor{mycoverage}{0.035}} \hfill
\figcap{\tiny \textcolor{mycard}{100}/\textcolor{mytex}{32}, \textcolor{mypsnr}{18.21}/\textcolor{mylpips}{0.176}/\textcolor{mycoverage}{0.034}} \hfill
\figcap{\tiny \textcolor{mycard}{100}/\textcolor{mytex}{64}, \textbf{\textcolor{mypsnr}{18.51}/\textcolor{mylpips}{0.165}/\textcolor{mycoverage}{0.029}}} \\
\includegraphics[trim = 20 0 40 10, clip, width=0.19\linewidth]{figs/taro/empty.jpg} \hfill
\includegraphics[trim = 20 0 40 10, clip, width=0.19\linewidth]{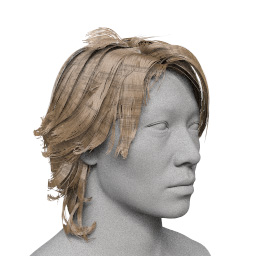} \hfill
\includegraphics[trim = 20 0 40 10, clip, width=0.19\linewidth]{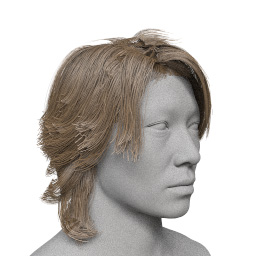} \hfill
\includegraphics[trim = 20 0 40 10, clip, width=0.19\linewidth]{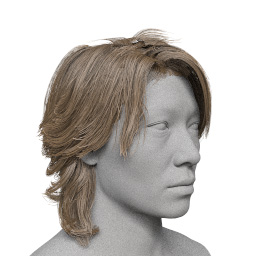} \hfill
\includegraphics[trim = 20 0 40 10, clip, width=0.19\linewidth]{figs/taro/ours_200_64.jpg}
\figcap{\tiny ~} \hfill
\figcap{\tiny \textcolor{mycard}{200}/\textcolor{mytex}{64}, \textcolor{mypsnr}{17.70}/\textcolor{mylpips}{0.200}/\textcolor{mycoverage}{0.032}} \hfill
\figcap{\tiny \textcolor{mycard}{200}/\textcolor{mytex}{16}, \textcolor{mypsnr}{18.99}/\textcolor{mylpips}{0.159}/\textcolor{mycoverage}{0.027}} \hfill
\figcap{\tiny \textcolor{mycard}{200}/\textcolor{mytex}{32}, \textcolor{mypsnr}{19.03}/\textcolor{mylpips}{0.150}/\textcolor{mycoverage}{0.025}} \hfill
\figcap{\tiny \textcolor{mycard}{200}/\textcolor{mytex}{64}, \textbf{\textcolor{mypsnr}{19.38}/\textcolor{mylpips}{0.138}/\textcolor{mycoverage}{0.021}}} \\
\caption{\label{fig:straight} Comparison with results from UE automatic hair card generator~\cite{HairCardGen}. $\textcolor{mycard}{\bullet}/\textcolor{mytex}{\bullet}, \textcolor{mypsnr}{\bullet}/\textcolor{mylpips}{\bullet}/\textcolor{mycoverage}{\bullet}$ indicate \textcolor{mycard}{the number of cards}, \textcolor{mytex}{the number of textures}, averaged \textcolor{mypsnr}{PSNR $\uparrow$}, \textcolor{mylpips}{LPIPS $\downarrow$}, and \textcolor{mycoverage}{coverage error $\downarrow$}, respectively.  }
\Description{}
\end{figure}

\paragraph{Ablation Study on Card \# and Texture \#}
\autoref{fig:straight} presents a comparison between our method and Unreal Engine's automatic hair card generator~\cite{HairCardGen} under varying numbers of cards and texture budgets. We evaluate results using 200, 100, and 50 cards, targeting low-resolution LoD hair generation. For textures, we test with 16 and 32 textures, following the industrial best practices for texture budgeting. As expected, visual similarity to the reference strand-based model decreases as the card and texture budgets are reduced. Therefore, the key to creating a good hair model lies in finding a balance between visual quality and geometric efficiency. Nevertheless, our method consistently achieves higher visual fidelity in all three metrics, even when using fewer textures and only 50 cards.

\begin{figure}[t!]
\centering
\newcommand{\figcap}[1]{\begin{minipage}{0.24\linewidth}\centering#1\end{minipage}}
\includegraphics[width=0.24\linewidth,trim =20 0 20 0, clip]{figs/wavy/strand.jpg} \hfill
\includegraphics[width=0.24\linewidth,trim =20 0 20 0, clip]{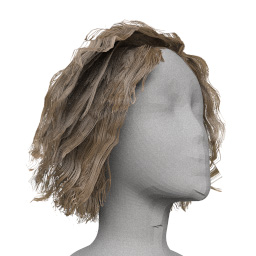} \hfill
\includegraphics[width=0.24\linewidth,trim =20 0 20 0, clip]{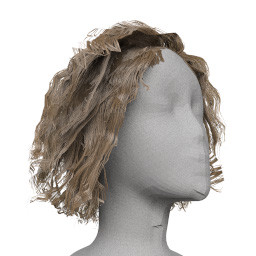} \hfill
\includegraphics[width=0.24\linewidth,trim =20 0 20 0, clip]{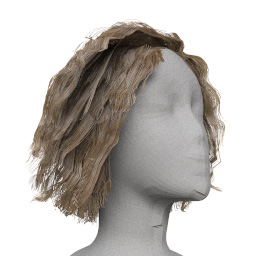} \\ \vspace{-1.5em}
\figcap{\small ~} \hfill
\figcap{\small \hspace{4.7em} \textbf{\textcolor{mypsnr}{19.83}}} \hfill
\figcap{\small \hspace{4.7em} \textcolor{mypsnr}{19.69}} \hfill
\figcap{\small \hspace{4.7em} \textcolor{mypsnr}{19.71}}\\
\figcap{\small Strands} \hfill
\figcap{\small Our total} \hfill
\figcap{\small w/o tangent loss} \hfill
\figcap{\small w/o depth loss } \\
\includegraphics[width=0.24\linewidth,trim =20 0 20 0, clip]{figs/taro/empty.jpg} \hfill
\includegraphics[width=0.24\linewidth,trim =20 0 20 0, clip]{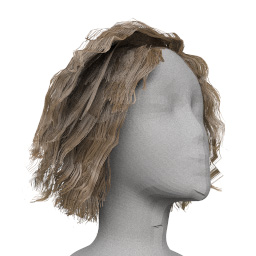} \hfill
\includegraphics[width=0.24\linewidth,trim =20 0 20 0, clip]{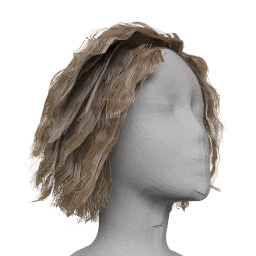} \hfill
\includegraphics[width=0.24\linewidth,trim =20 0 20 0, clip]{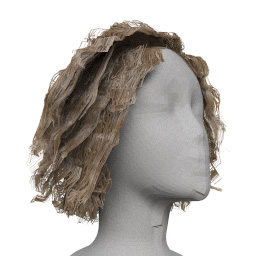}\\  \vspace{-1.5em}
\figcap{\small ~} \hfill
\figcap{\small  \hspace{4.7em} \textcolor{mypsnr}{19.82}} \hfill
\figcap{\small  \hspace{4.7em} \textcolor{mypsnr}{19.67}} \hfill
\figcap{\small  \hspace{4.7em} \textcolor{mypsnr}{19.53}}\\
\figcap{\small ~} \hfill
\figcap{\small w/o collision loss} \hfill
\figcap{\small w/o dice loss} \hfill
\figcap{\small w/o match loss} \\
\caption{\label{fig:ablation}Ablation study on individual loss components. Removing any of the proposed loss terms either degrades the visual quality or increases the deviation from the strand-based model.  $\textcolor{mypsnr}{\bullet}$ indicates \textcolor{mypsnr}{PSNR $\uparrow$}.
} 
\Description{}
\end{figure}

\paragraph{Ablation Study on Losses}
We further conduct an ablation study on all loss and regularization terms on a straight hairstyle from MetaHuman~\cite{unrealengine} with 100 hair cards and 32 textures. As shown in~\autoref{fig:ablation}, each component is necessary to achieve consistent and accurate results. Unfortunately, due to the subpixel nature of hair strands and their complicated geometries, traditional image-based metrics such as MSE and PSNR cannot fully capture the perceptual quality of hair. For instance, omitting tangent loss often results in unnatural zigzag strands. Similarly, allowing cards to intersect with the head may seem minor in visual metrics, but it causes noticeable artifacts during simulation. 
For the rest of the experiments, we adopt two sets of loss weights for different hair types. For straight hair, we use $\lambda^\text{tangent} = 10$, $\lambda^\text{depth} = 10$, $\lambda^\text{dice} = 5$, $\lambda^\text{match} = 3$, and $\lambda^\text{collision} = 1 \times 10^5$. For curly hair, we use a different configuration as $\lambda^\text{tangent} = 5$, $\lambda^\text{depth} = 15$, $\lambda^\text{dice} = 3$, $\lambda^\text{match} = 3$, and $\lambda^\text{collision} = 1 \times 10^5$, to encourage more curl strands. Note that, due to the inherent geometric differences between straight and curly hair, we adopt two sets of configurations that work well across all examples. We believe this choice does not compromise the generalizability of our method.

\begin{figure}[t!]
\newcommand{\figcap}[1]{\begin{minipage}{0.24\linewidth}\centering#1\end{minipage}}
\centering
\figcap{\small Strands \\~} \hfill
\figcap{\small UE cards \\~} \hfill
\figcap{\small Our cards without cap}\hfill
\figcap{\small Our cards \\ with cap}\\
\includegraphics[trim =80 0 80 75, clip, width=0.24\linewidth]{figs/short/strand.jpg} \hfill
\includegraphics[trim =80 0 80 75, clip, width=0.24\linewidth]{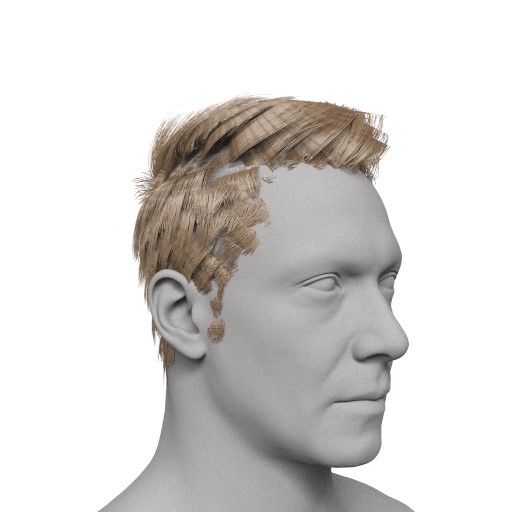}\hfill
\includegraphics[trim =80 0 80 75, clip, width=0.24\linewidth]{figs/short/our_card.jpg} \hfill
\includegraphics[trim =80 0 80 75, clip, width=0.24\linewidth]{figs/short/our_card_cap.jpg} \\
\figcap{\small Short} \hfill
\figcap{\small \textcolor{mypsnr}{22.34}/\textcolor{mylpips}{0.112}/\textcolor{mycoverage}{0.014}} \hfill
\figcap{\small {\textcolor{mypsnr}{23.34}/\textcolor{mylpips}{0.100}/\textcolor{mycoverage}{0.011}}} \hfill
\figcap{\small \textbf{\textcolor{mypsnr}{23.51}/\textcolor{mylpips}{0.100}/\textcolor{mycoverage}{0.010}}} \\
\includegraphics[trim =30 10 30 10, clip, width=0.24\linewidth]{figs/bangs/strand.jpg} \hfill
\includegraphics[trim =30 10 30 10, clip, width=0.24\linewidth]{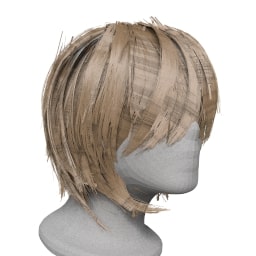}\hfill
\includegraphics[trim =30 10 30 10, clip, width=0.24\linewidth]{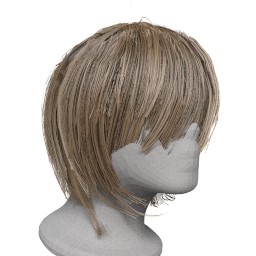} \hfill
\includegraphics[trim =30 10 30 10, clip, width=0.24\linewidth]{figs/bangs/ours.jpg} \\
\figcap{\small Bangs} \hfill
\figcap{\small \textcolor{mypsnr}{18.09}/\textcolor{mylpips}{0.218}/\textcolor{mycoverage}{0.032}} \hfill
\figcap{\small {\textcolor{mypsnr}{20.38}/\textcolor{mylpips}{0.154}/\textcolor{mycoverage}{0.019}}} \hfill
\figcap{\small \textbf{\textcolor{mypsnr}{20.47}/\textcolor{mylpips}{0.154}/\textcolor{mycoverage}{0.018}}} \\
\includegraphics[trim =30 10 35 10, clip, width=0.24\linewidth]{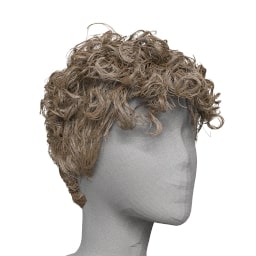} \hfill
\includegraphics[trim =30 10 35 10, clip, width=0.24\linewidth]{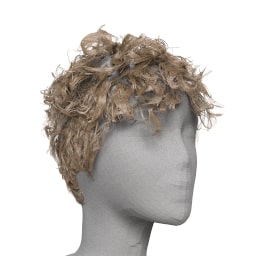}\hfill
\includegraphics[trim =30 10 35 10, clip, width=0.24\linewidth]{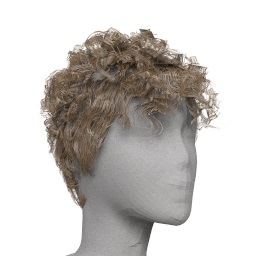} \hfill
\includegraphics[trim =30 10 35 10, clip, width=0.24\linewidth]{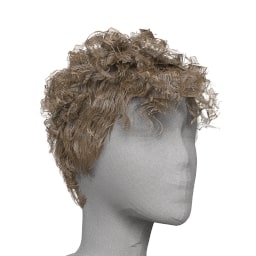} \\
\figcap{\small Curly} \hfill
\figcap{\small \textcolor{mypsnr}{18.54}/\textcolor{mylpips}{0.153}/\textcolor{mycoverage}{0.038}} \hfill
{\small {\textcolor{mypsnr}{20.75}/\textcolor{mylpips}{0.138}/\textcolor{mycoverage}{0.018}}}\hfill
\figcap{\small \textbf{\textcolor{mypsnr}{20.89}/\textcolor{mylpips}{0.138}/\textcolor{mycoverage}{0.016}}} \\
\includegraphics[trim =30 30 30 0, clip, width=0.24\linewidth]{figs/blowout/strand.jpg} \hfill
\includegraphics[trim =30 30 30 0, clip, width=0.24\linewidth]{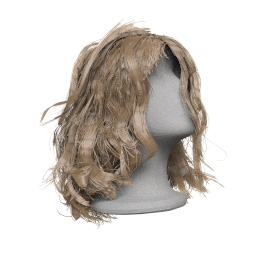} \hfill
\includegraphics[trim =30 30 30 0, clip, width=0.24\linewidth]{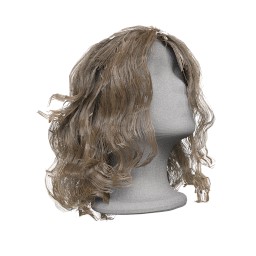}\hfill
\includegraphics[trim =30 30 30 0, clip, width=0.24\linewidth]{figs/blowout/ours_result_wavy.jpg}\\
\figcap{\small Blowout} \hfill
\figcap{\small \textcolor{mypsnr}{17.16}/\textcolor{mylpips}{0.200}/\textcolor{mycoverage}{0.038}} \hfill
\figcap{\small \textcolor{mypsnr}{20.40}/\textcolor{mylpips}{0.162}/\textcolor{mycoverage}{0.017}}\hfill
\figcap{\small \textbf{\textcolor{mypsnr}{20.44}/\textcolor{mylpips}{0.162}/\textcolor{mycoverage}{0.017}}} \\
\includegraphics[trim =10 0 10 0, clip, width=0.24\linewidth]{figs/wavy/strand.jpg} \hfill
\includegraphics[trim =10 0 10 0, clip, width=0.24\linewidth]{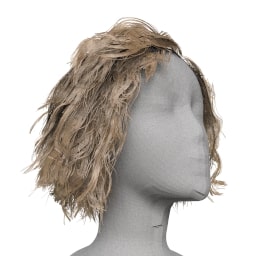} \hfill
\includegraphics[trim =10 0 10 0, clip, width=0.24\linewidth]{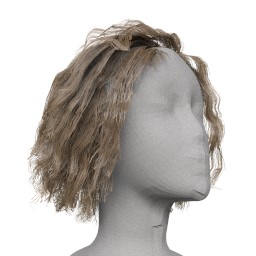}\hfill
\includegraphics[trim =10 0 10 0, clip, width=0.24\linewidth]{figs/wavy/ours_scale3.jpg} \\ 
\figcap{\small Wavy} \hfill
\figcap{\small \textcolor{mypsnr}{16.82}/\textcolor{mylpips}{0.228}/\textcolor{mycoverage}{0.041}} \hfill
\figcap{\small {\textcolor{mypsnr}{19.89}/\textcolor{mylpips}{0.166}/\textcolor{mycoverage}{0.018}}} \hfill
\figcap{\small \textbf{\textcolor{mypsnr}{19.89}/\textcolor{mylpips}{0.166}/\textcolor{mycoverage}{0.016}}} \\
\caption{\label{fig:gallary} Our results outperform UE cards over a variety of hair styles, including short, bangs, wavy, curly, and blowout, regardless of whether with a hair cap or not.
}
\Description{}
\end{figure}

\paragraph{Ablation Study on Hair Caps} 
Our results outperform Unreal Engine's cards over a variety of hair styles, including short, bangs, wavy, curly, and blowout, as shown in~\autoref{fig:gallary}. The hair cap primarily improves scalp coverage and hairline, but does not affect overall hair silhouette. As for the geometry, hair cards inherently involve a trade-off between visual fidelity and geometric accuracy. This work primarily aims to reproduce the overall appearance within a constrained geometric budget. Hair geometry that is barely visible contributes little to appearance, so precise geometric accuracy in those regions is less critical. For this reason, our pipeline is appearance-driven, leveraging differentiable rendering to optimize hair cards for visual similarity rather than exact geometry reproduction.

\begin{figure}[t!]
\newcommand{\figcap}[1]{\begin{minipage}{0.325\linewidth}\centering#1\end{minipage}}
\centering
\figcap{\small Strands} \hfill
\figcap{\small Init w/ UE cards} \hfill
\figcap{\small Ours}\\
\includegraphics[trim =20 0 30 10, clip, width=0.325\linewidth]{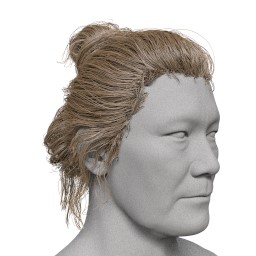} \hfill
\includegraphics[trim =20 0 30 10, clip, width=0.325\linewidth]{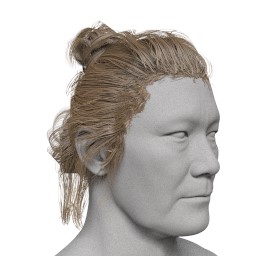} \hfill
\includegraphics[trim =20 0 30 10, clip, width=0.325\linewidth]{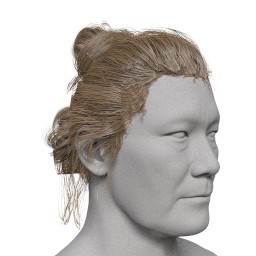}\\ 
\figcap{\small~} \hfill
\figcap{\small \textcolor{mypsnr}{17.27}/\textcolor{mylpips}{0.232}/\textcolor{mycoverage}{0.046}} \hfill
\figcap{\small \textbf{\textcolor{mypsnr}{19.30}/\textcolor{mylpips}{0.158}/\textcolor{mycoverage}{0.027}}} \\
\caption{\label{fig:ablation_geo} Ablation study on card initialization. Using UE cards as initialization degrades the visual quality. } 
\Description{}
\end{figure}

\begin{figure}[t!]
\newcommand{\figcap}[1]{\begin{minipage}{0.24\linewidth}\centering#1\end{minipage}}
\centering
\figcap{\small Strands} \hfill
\figcap{\small Manual cards} \hfill
\figcap{\small UE auto cards} \hfill
\figcap{\small Our cards} \\
\includegraphics[trim =30 0 30 0, clip, width=0.24\linewidth]{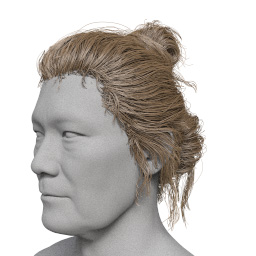} \hfill
\includegraphics[trim =60 0 60 0, clip, width=0.24\linewidth]{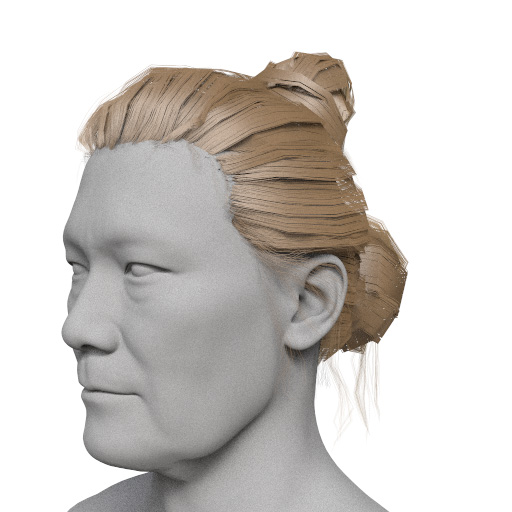} \hfill
\includegraphics[trim =30 0 30 0, clip, width=0.24\linewidth]{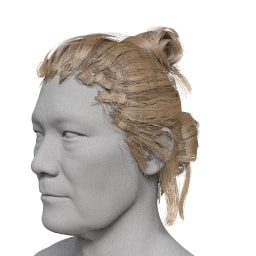} \hfill
\includegraphics[trim =30 0 30 0, clip, width=0.24\linewidth]{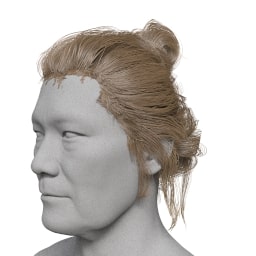} \\ 
\includegraphics[trim =30 0 30 0, clip, width=0.24\linewidth]{figs/aoi/strand_90.jpg} \hfill
\includegraphics[trim =60 0 60 0, clip, width=0.24\linewidth]{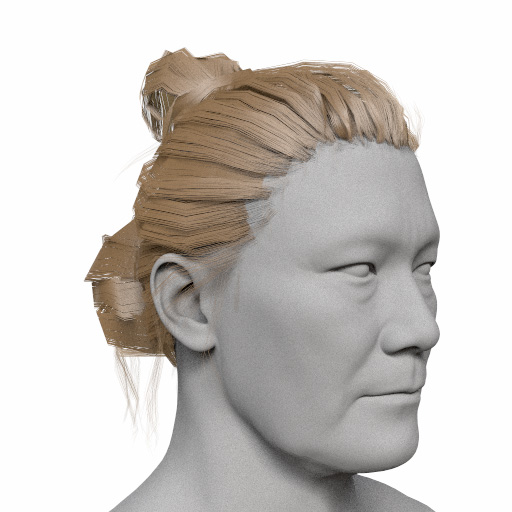} \hfill
\includegraphics[trim =30 0 30 0, clip, width=0.24\linewidth]{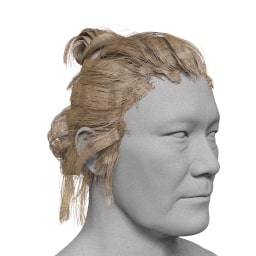} \hfill
\includegraphics[trim =30 0 30 0, clip, width=0.24\linewidth]{figs/aoi/ours_351_90_pipeline.jpg} \\
\includegraphics[trim =30 0 30 0, clip, width=0.24\linewidth]{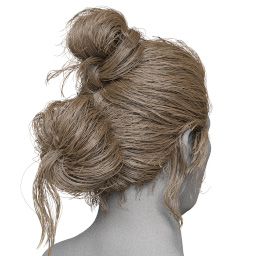} \hfill
\includegraphics[trim =60 0 60 0, clip, width=0.24\linewidth]{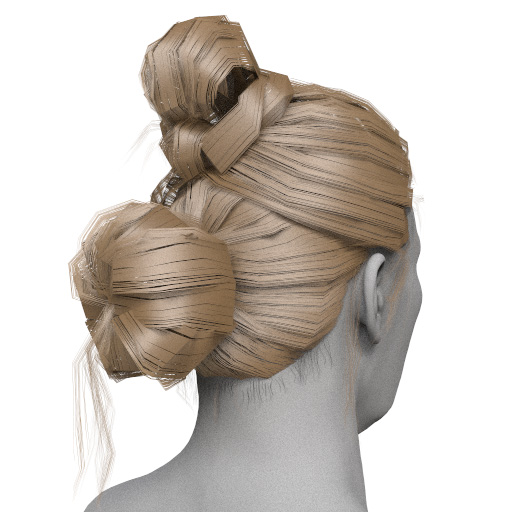} \hfill
\includegraphics[trim =30 0 30 0, clip, width=0.24\linewidth]{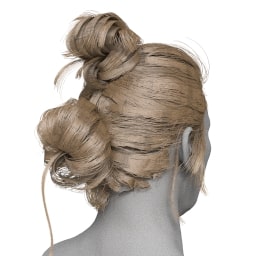} \hfill
\includegraphics[trim =30 0 30 0, clip, width=0.24\linewidth]{figs/aoi/ours_351_180_pipeline.jpg} \\ 
\figcap{\small~} \hfill
\figcap{\small \textcolor{mypsnr}{17.71}/\textcolor{mylpips}{0.279}/\textcolor{mycoverage}{0.044}} \hfill
\figcap{\small \textcolor{mypsnr}{17.78}/\textcolor{mylpips}{0.209}/\textcolor{mycoverage}{0.038}} \hfill
\figcap{\small \textbf{\textcolor{mypsnr}{19.99}/\textcolor{mylpips}{0.140}/\textcolor{mycoverage}{0.021}}} \\
\caption{\label{fig:bun} Comparison with Unreal Engine's auto-generated and manually crafted cards. Our method outperforms the other two in all metrics with the same 351 cards and 32 textures.
}
\Description{}
\end{figure}

\begin{figure}[t!]
\newcommand{\figcap}[1]{\begin{minipage}{0.19\linewidth}\centering#1\end{minipage}}
\centering
\includegraphics[trim =125 0 225 150, clip, width=0.16\linewidth]{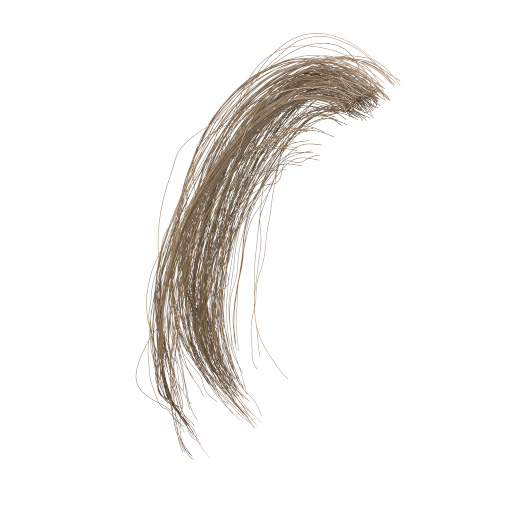} \hfill
\includegraphics[trim =125 0 225 150, clip, width=0.16\linewidth]{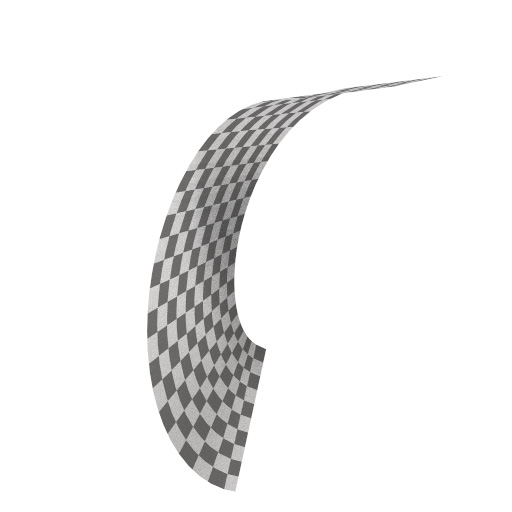} \hfill
\includegraphics[trim =125 0 225 150, clip, width=0.16\linewidth]{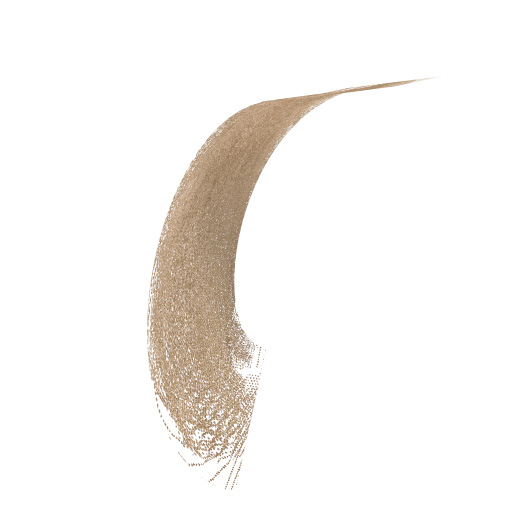} \hfill
\includegraphics[trim =125 0 225 150, clip, width=0.16\linewidth]{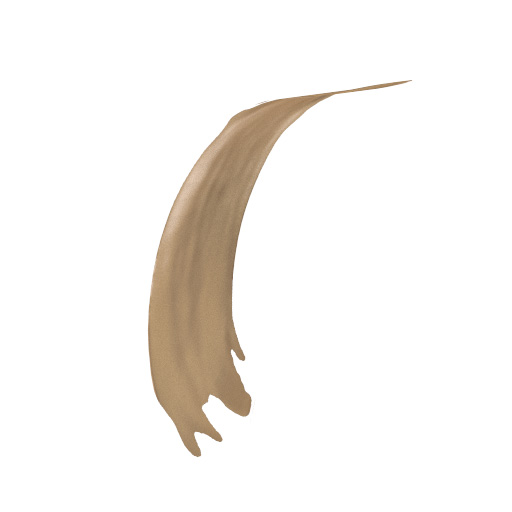} \hfill
\includegraphics[trim =125 0 225 150, clip, width=0.16\linewidth]{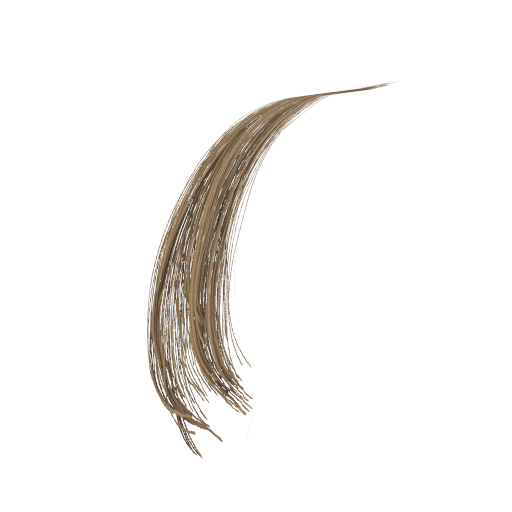}\\
\figcap{\small Strands} \hfill
\figcap{\small Input card} \hfill
\figcap{\small Tex-based \\ w/o smooth} \hfill
\figcap{\small Tex-based \\ w/ smooth} \hfill
\figcap{\small Ours}
\caption{\label{fig:opttex} Given input card and hair views from eight directions, directly optimizing texture leads to noisy results, while adding smoothness loss eliminates high-frequency hair structure. Our strand-based optimization avoids the above issues effectively.}
\Description{}
\end{figure}

\begin{figure}[t!]
\newcommand{\figcap}[1]{\begin{minipage}{0.325\linewidth}\centering#1\end{minipage}}
\centering
\figcap{\small Strands} \hfill
\figcap{\small UE cards} \hfill
\figcap{\small Our cards} \\
\includegraphics[trim =50 20 35 15, clip, width=0.325\linewidth]{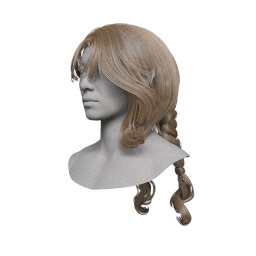} \hfill
\includegraphics[trim =50 20 35 15, clip, width=0.325\linewidth]{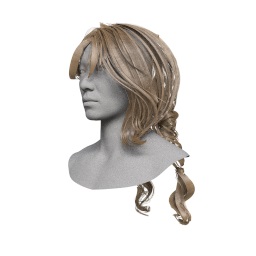} \hfill
\includegraphics[trim =50 20 35 15, clip, width=0.325\linewidth]{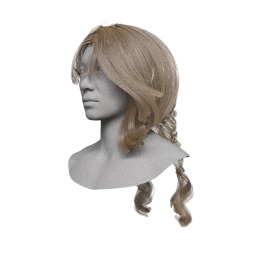}\\ 
\figcap{\small~} \hfill
\figcap{\small \textcolor{mypsnr}{23.25}/\textcolor{mylpips}{0.086}/\textcolor{mycoverage}{0.010}} \hfill
\figcap{\small \textbf{\textcolor{mypsnr}{24.90}/\textcolor{mylpips}{0.077}/\textcolor{mycoverage}{0.007}}}
\caption{\label{fig:braid} Our method applies equally well to complex structures such as knots and braids. }
\Description{}
\end{figure}

\begin{figure}[t!]
\newcommand{\figcap}[1]{\begin{minipage}{0.325\linewidth}\centering#1\end{minipage}}
\centering
\figcap{\small Strands} \hfill
\figcap{\small UE cards} \hfill
\figcap{\small Our cards} \\
\includegraphics[trim =20 0 10 25, clip, width=0.325\linewidth]{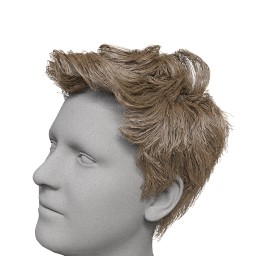} \hfill
\includegraphics[trim =20 0 10 25, clip, width=0.325\linewidth]{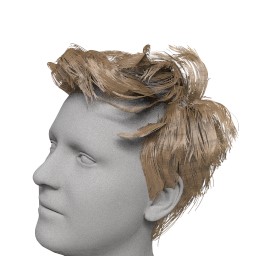} \hfill
\includegraphics[trim =20 0 10 25, clip, width=0.325\linewidth]{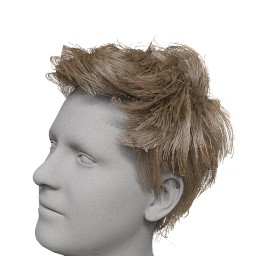}\\ 
\figcap{\small Fringe} \hfill
\figcap{\small \textcolor{mypsnr}{18.47}/\textcolor{mylpips}{0.166}/\textcolor{mycoverage}{0.036}} \hfill
\figcap{\small \textbf{\textcolor{mypsnr}{19.24}/\textcolor{mylpips}{0.158}/\textcolor{mycoverage}{0.029}}}
\caption{\label{fig:multiview} Hair card extraction for strand-based hairs from multi-view reconstruction~\cite{zakharov2024human}. }
\Description{}
\end{figure}

\paragraph{Ablation Study on Card Initialization}
We conduct an ablation study on card orientation optimization on Bun hairstyle from MetaHuman~\cite{unrealengine} with 200 hair cards and 32 textures. As shown in~\autoref{fig:ablation_geo}, using cards generated from the UE automatic hair card generator as the initial shape for our optimization results in a degradation of visual quality, while our card geometry initialization leads to better optimized results. 

\paragraph{Comparison with UE Auto-generated and Manual-crafted Cards} 
We compare our results with the output of UE built-in automatic hair card generation tool~\cite{HairCardGen} and artist-crafted hair cards from MetaHuman~\cite{unrealengine}. We compare with the lowest LoD hair card model from MetaHuman, which only contains 351 cards and 20 textures. As shown in~\autoref{fig:bun}, using the same set of 351 hair cards and 20 textures, our method achieves better visual fidelity to the input strand-based model than both Unreal Engine auto-generated and manual-crafted cards. While our higher PSNR partly benefits from more accurate AO textures, as we preserve the displacement during optimization, the lower LPIPS and coverage errors clearly demonstrate that our results more faithfully preserve hair occupancy and shape.

\begin{figure}[t!]
\newcommand{\figcap}[1]{\begin{minipage}{0.49\linewidth}\centering#1\end{minipage}}
\centering
\includegraphics[trim =180 150 180 100, clip, width=\linewidth]{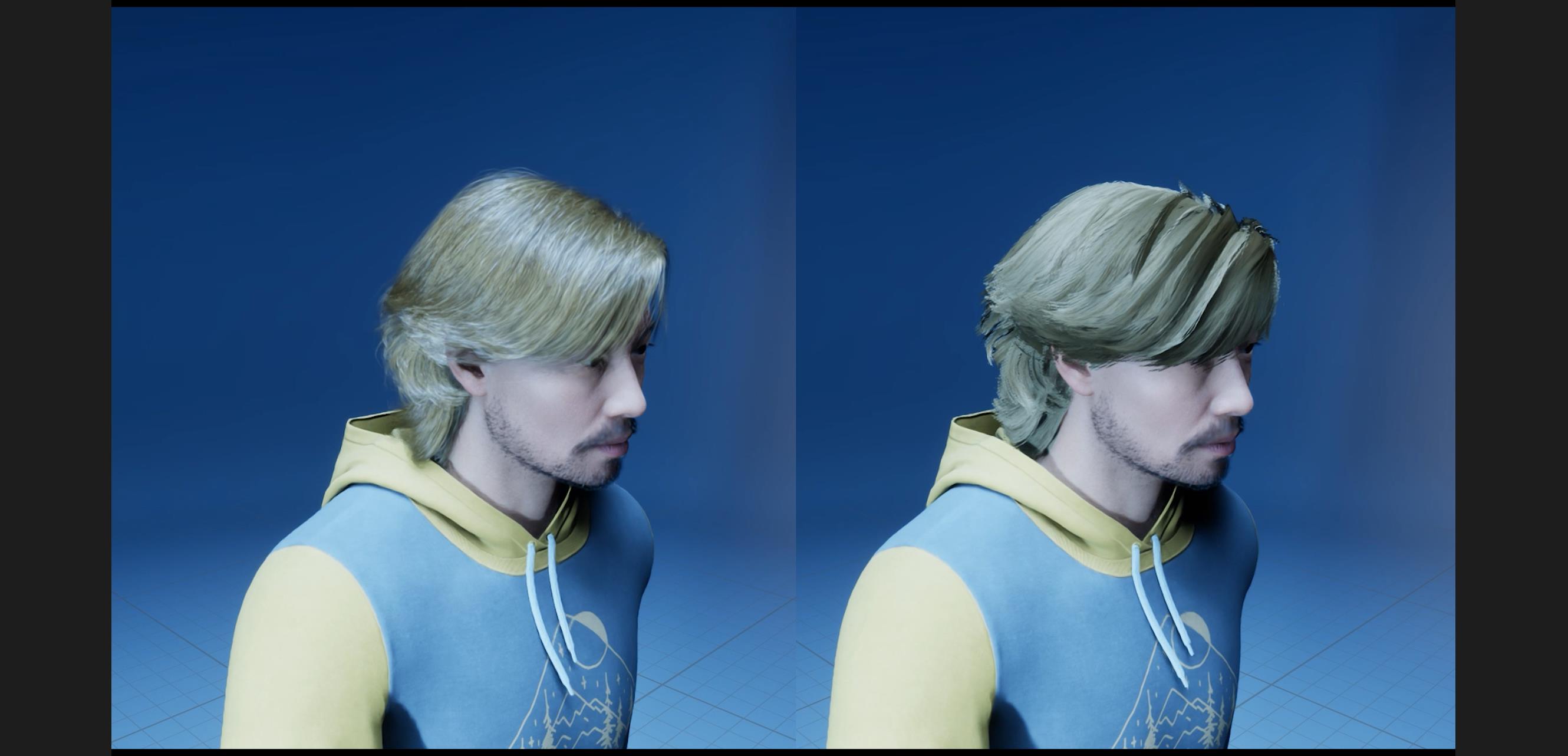} \\
\includegraphics[trim =180 230 180 20, clip, width=\linewidth]{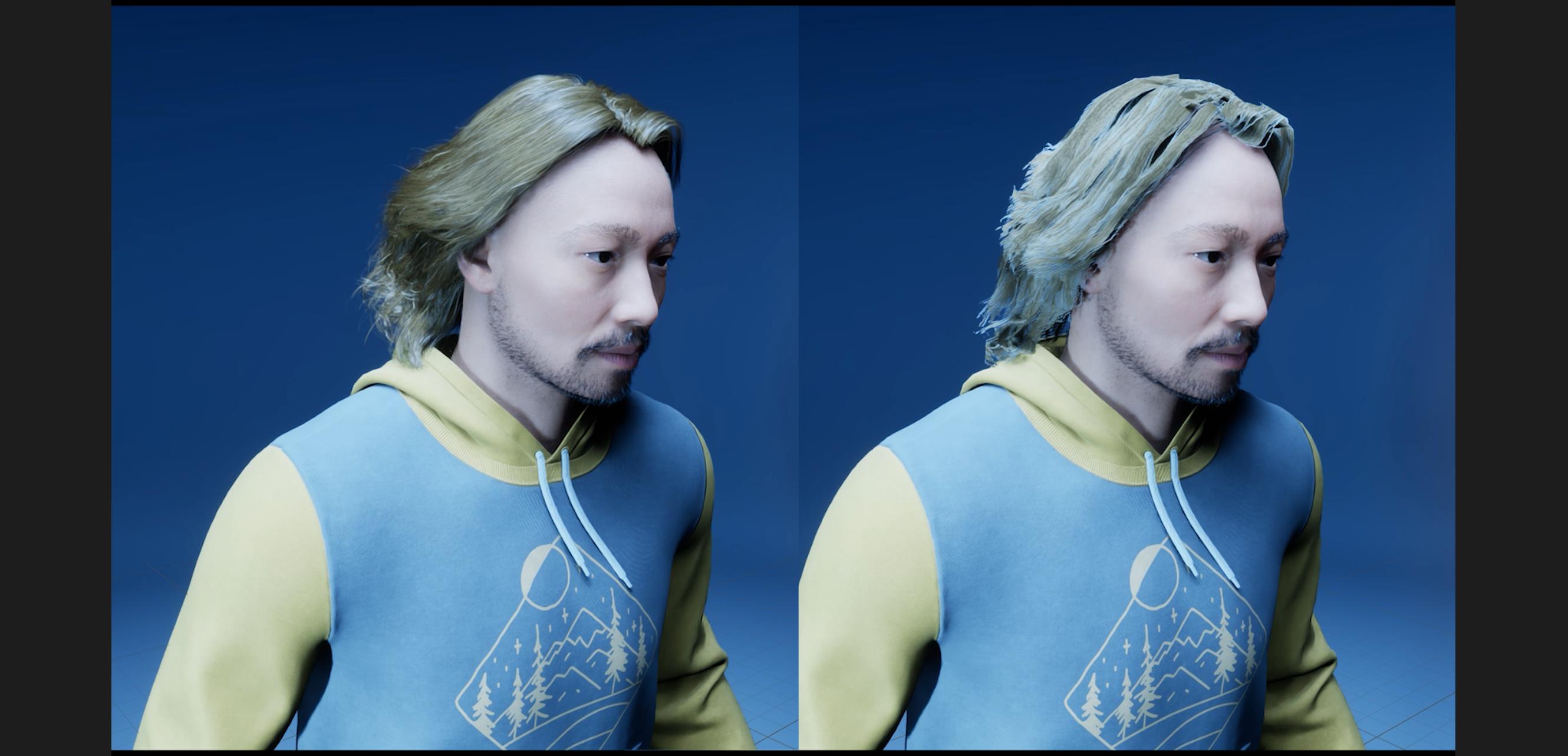} \\
\includegraphics[trim =180 150 180 100, clip, width=\linewidth]{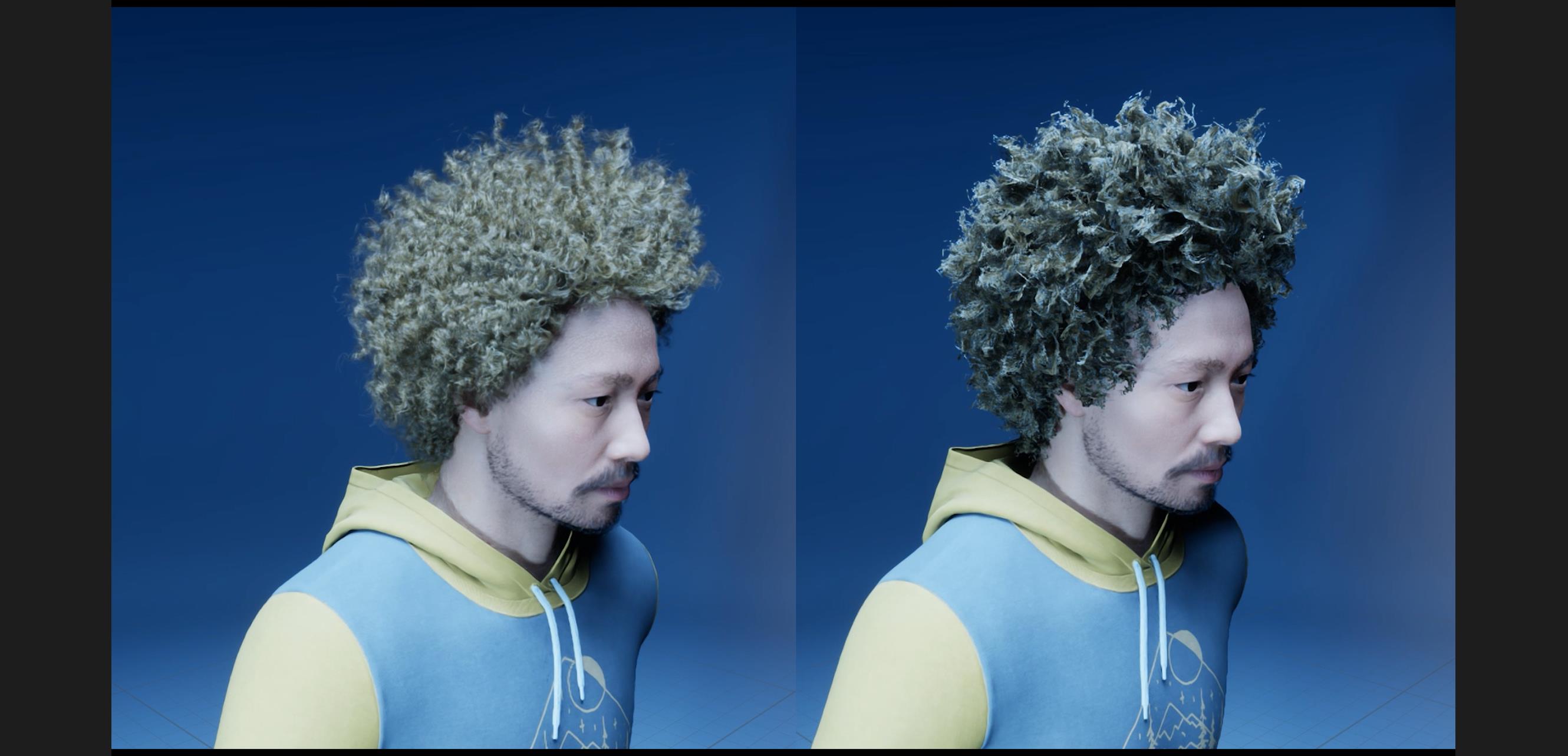} \\
\figcap{\small Strands} \hfill
\figcap{\small Our cards} \\
\caption{\label{fig:demo} Comparison between simulating and rendering the input strand-based model and our output hair card model in Unreal Engine~\cite{unrealengine}.}
\Description{}
\end{figure}

\paragraph{Comparison with Optimizing Texture Directly}
We validate the necessity of our explicit strand representation by comparing it with a conventional differentiable rendering pipeline that directly optimizes hair card textures using Nvdiffrast~\cite{Laine2020diffrast}. In this experiment, we provide a fixed hair card geometry along with tangent images from eight view directions. We then use Nvdiffrast to optimize the texture mapped to the card so that its rendered appearance matches the given tangent images as closely as possible from those views. Although Nvdiffrast can produce hair textures with distinct shape features when optimizing for a single view direction, extending the optimization to multiple views leads to noisy and inconsistent results. Introducing a smoothness loss can eliminate noise at the cost of losing high-frequency details, as shown in~\autoref{fig:opttex}. More importantly, this texture-based approach lacks the spatial structure for computing ambient occlusion accurately, as hair-to-hair spatial relationships are not preserved during texture optimization.

\begin{figure*}[t!]
\newcommand{\figcap}[1]{\begin{minipage}{0.19\linewidth}\centering#1\end{minipage}}
\centering
\figcap{\small Strands} \hfill
\figcap{\small Ours, 20 cards} \hfill
\figcap{\small Ours, 50 cards} \hfill
\figcap{\small Ours, 100 cards}\hfill
\figcap{\small Ours, 200 cards} \\
\includegraphics[trim =20 0 30 10, clip, width=0.19\linewidth]{figs/taro/strand.jpg} \hfill
\includegraphics[trim =20 0 30 10, clip, width=0.19\linewidth]{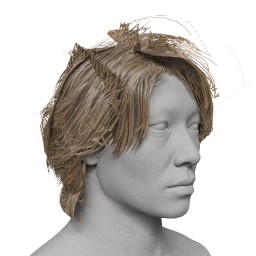} \hfill
\includegraphics[trim =20 0 30 10, clip, width=0.19\linewidth]{figs/taro/ours_50_16.jpg}\hfill
\includegraphics[trim =20 0 30 10, clip, width=0.19\linewidth]{figs/taro/ours_100_16.jpg} \hfill
\includegraphics[trim =20 0 30 10, clip, width=0.19\linewidth]{figs/taro/ours_200_16.jpg}\\ 
\figcap{\small ~} \hfill
\figcap{\small \textcolor{mycard}{20}/\textcolor{mytex}{16}, \textcolor{mypsnr}{17.65}/\textcolor{mylpips}{0.245}/\textcolor{mycoverage}{0.041}} \hfill
\figcap{\small \textcolor{mycard}{50}/\textcolor{mytex}{16}, \textcolor{mypsnr}{17.66}/\textcolor{mylpips}{0.203}/\textcolor{mycoverage}{0.045} }\hfill
\figcap{\small \textcolor{mycard}{100}/\textcolor{mytex}{16}, \textcolor{mypsnr}{18.03}/\textcolor{mylpips}{0.192}/\textcolor{mycoverage}{0.035}}
\figcap{\small \textcolor{mycard}{200}/\textcolor{mytex}{16}, \textcolor{mypsnr}{18.99}/\textcolor{mylpips}{0.159}/\textcolor{mycoverage}{0.027}} \hfill\\
\caption{\label{fig:lowcards}  Using too few hair cards, for example, only 20, fails to adequately represent the hair structure, even for relatively simple straight styles. As the number of cards increases, the overall fidelity of the hair representation improves. $\textcolor{mycard}{\bullet}/\textcolor{mytex}{\bullet}, \textcolor{mypsnr}{\bullet}/\textcolor{mylpips}{\bullet}/\textcolor{mycoverage}{\bullet}$ indicate \textcolor{mycard}{the number of cards}, \textcolor{mytex}{the number of textures}, averaged \textcolor{mypsnr}{PSNR $\uparrow$}, \textcolor{mylpips}{LPIPS $\downarrow$}, and \textcolor{mycoverage}{coverage error $\downarrow$}, respectively. } 
\Description{}
\end{figure*}

\paragraph{Braid Styles}
\autoref{fig:pipeline} demonstrates that buns can be grouped based on their geometric and spatial similarity, a strategy that applies equally well to complex structures such as knots and braids. \autoref{fig:braid} shows another example of braided hairstyles, where our method can produce hair cards that reproduce the braided style with a small number of hair cards.

\paragraph{Strand-based Hairs from Multi-view Reconstruction}
To further validate the scope of application of our method, we test our method on the strand model generated using the multi-view reconstruction~\cite{zakharov2024human}. As shown in ~\autoref{fig:multiview}, our method can automatically extract hair cards from a reconstructed strand-based fringe hairstyle.

\paragraph{Comparison with Strand Model in Game Engine}
\autoref{fig:demo} demonstrates that our hair cards can be used directly in Unreal Engine with real-time rendering and simulation. Specifically, we generate one guide hair per card for simulation and drive the deformation of the hair cards using linear blend skinning (LBS) as mentioned in~\cite{Hsu2024}. The strand-based model with approximately 50,000 strands requires 2.4 ms per frame for rendering, whereas our card model only takes 0.8 ms for 512 hair cards. Please refer to the supplemental video.

\paragraph{Cards Extraction Performance}
Given input hair with 40K strands with $n^s=32$ samples, the average process time of our card generation pipeline for cards with 100 cards and 32 textures is about 46 minutes, where joint optimization takes about $42\%$ of the computation time. For a larger configuration of 400 cards with the same number of textures, the total runtime increases to approximately 1 hour, with joint optimization comprising roughly one-third of the computation time. In both cases, the optimization process converges within roughly $200$ epochs. Memory cost is about 7 GB for $100$ cards and 10 GB for $400$ cards, respectively. 

%% file: conclusion.tex
\begin{figure*}[t!]
\newcommand{\figcap}[1]{\begin{minipage}{0.196\linewidth}\centering#1\end{minipage}}
\centering
\figcap{\small Strands} \hfill
\figcap{\small Unreal Engine's 800 cards} \hfill
\figcap{\small Ours, 50$\times$2 cards} \hfill
\figcap{\small Ours, 400$\times$2 cards}\hfill
\figcap{\small Ours, 20000$\times$2 cards} \\
\includegraphics[trim =20 0 20 10, clip, width=0.19\linewidth]{figs/coily/strand.jpg} \hfill
\includegraphics[trim =20 0 20 10, clip, width=0.19\linewidth]{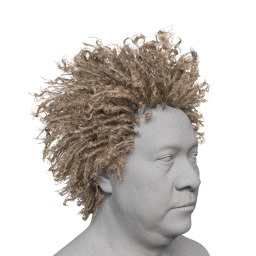} \hfill
\includegraphics[trim =20 0 20 10, clip, width=0.19\linewidth]{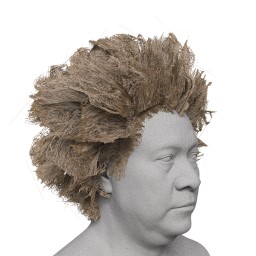}\hfill
\includegraphics[trim =20 0 20 10, clip, width=0.19\linewidth]{figs/coily/ours.jpg} \hfill
\includegraphics[trim =20 0 20 10, clip, width=0.19\linewidth]{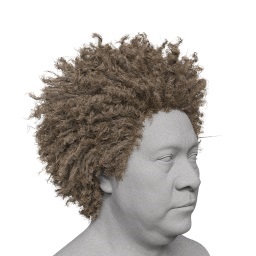}\\ 
\includegraphics[trim =20 0 20 10, clip, width=0.19\linewidth]{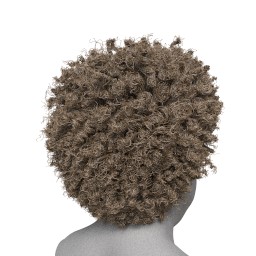} \hfill
\includegraphics[trim =20 0 20 10, clip, width=0.19\linewidth]{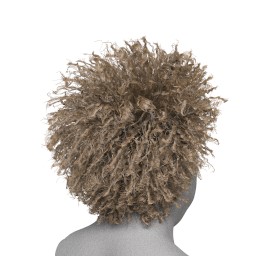} \hfill
\includegraphics[trim =20 0 20 10, clip, width=0.19\linewidth]{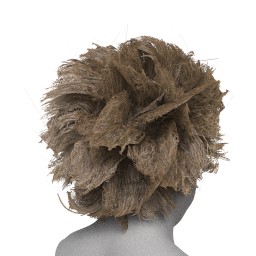}\hfill
\includegraphics[trim =20 0 20 10, clip, width=0.19\linewidth]{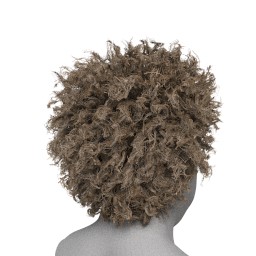} \hfill
\includegraphics[trim =20 0 20 10, clip, width=0.19\linewidth]{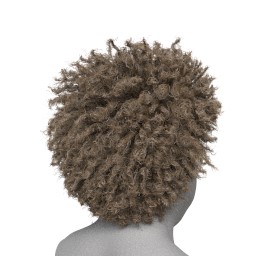}\\
\figcap{\small ~} \hfill
\figcap{\small \textcolor{mycard}{800}/\textcolor{mytex}{32}, \textcolor{mypsnr}{17.90}/\textcolor{mylpips}{0.174}/\textcolor{mycoverage}{0.032}} \hfill
\figcap{\small \textcolor{mycard}{50$\times$2}/\textcolor{mytex}{32}, \textcolor{mypsnr}{18.13}/\textcolor{mylpips}{0.253}/\textcolor{mycoverage}{0.028}} \hfill
\figcap{\small \textcolor{mycard}{400$\times$2}/\textcolor{mytex}{32}, \textbf{\textcolor{mypsnr}{19.09}/\textcolor{mylpips}{0.131}/\textcolor{mycoverage}{0.022}}}\hfill
\figcap{\small \textcolor{mycard}{20000$\times$2}/\textcolor{mytex}{32}, \textbf{\textcolor{mypsnr}{21.45}/\textcolor{mylpips}{0.093}/\textcolor{mycoverage}{0.011}}} \\
\caption{\label{fig:coily} While our method can reproduce the appearance of coily hair with a relatively high number of cards (e.g., 20,000$\times$2), it becomes challenging to preserve fidelity when the number of cards is limited. $\textcolor{mycard}{\bullet}/\textcolor{mytex}{\bullet}, \textcolor{mypsnr}{\bullet}/\textcolor{mylpips}{\bullet}/\textcolor{mycoverage}{\bullet}$ indicate \textcolor{mycard}{the number of cards}, \textcolor{mytex}{the number of textures}, averaged \textcolor{mypsnr}{PSNR $\uparrow$}, \textcolor{mylpips}{LPIPS $\downarrow$}, and \textcolor{mycoverage}{coverage error $\downarrow$}, respectively. }
\Description{}
\end{figure*}

\section{Conclusion}

We have presented a fully automated pipeline for converting strand-based hair models into efficient and visually compelling hair card representations. By leveraging a differentiable rendering framework, our method first clusters the hair strands to initialize card geometry and then clusters the hair textures to share, thereby reducing memory cost. Finally, we conduct joint optimization over both card geometry and textures. A key contribution of our approach is the introduction of 2D curve-based texture encoding, which offers a resolution-independent representation that effectively captures fine strand details while remaining compatible with differentiable rendering. Our method supports a wide range of hairstyles and lengths, introducing mechanisms such as hair caps and crossed card generation to handle visually complex hair types, including short and coily styles. Additionally, the ability to share textures across LoDs makes our approach well-suited for real-time applications where memory efficiency and performance are critical.

\paragraph{Limitations}
Although our pipeline demonstrates strong performance and visual fidelity across a range of hairstyles, several limitations remain. First, optimization is conducted in multiple substeps rather than as a fully end-to-end differentiable pipeline, which may limit global consistency. Given the already large number of variables to optimize (approximately 500K variables for the afro example), increasing the number of strand control points would indeed improve the representation capability of strands. However, this also enlarges the solution space, making optimization more prone to getting trapped in local minima. Second, the loss functions primarily rely on visible attributes such as tangent alignment, depth, and mask coverage. However, for complex hairstyles, especially those with dense or layered structures, interior strand arrangements play an important visual role and are not explicitly enforced. Besides, hair cards, as a low-cost representation, have an inherent limitation in that it is mostly for static and moderate animation in games. Under large deformation, it can cause occlusion artifacts. Third, our method assumes a uniform hair color throughout the entire hairstyle, which may not generalize well to stylized or multicolored hair. Finally, hair type with highly curved geometry, such as coily hair, remains a challenge. Theoretically, increasing the number of hair clusters and card geometries makes it easier to represent complex hairstyles. Ultimately, if we assign one card per strand, our pipeline would always succeed. However, our goal is to strike a balance between geometry and appearance. \autoref{fig:lowcards} shows that using too few hair cards, for example, only 20, fails to adequately represent the hair structure, even for relatively simple straight styles. As the number of cards increases, the overall fidelity of the hair representation improves. Similarly, \autoref{fig:coily} demonstrates that our method struggles to capture complex coily structures when both the number of hair cards and control points are limited. However, with a sufficiently large number of cards (e.g., 20,000$\times$2), even intricate coily patterns can be represented with fine detail.
